\documentclass{aa}  

\usepackage{graphicx}
\usepackage{txfonts}
\usepackage{multirow}
\usepackage{amssymb}
\usepackage{units}
\usepackage{float}
\usepackage{natbib}
\bibpunct{(}{)}{;}{a}{}{,}

\usepackage[dvipsnames]{xcolor}
\usepackage{color, colortbl}
\usepackage{xspace}
\usepackage{lscape}
\usepackage{url}
\usepackage{arydshln}

\usepackage[hyperfootnotes=true, linktocpage=true, breaklinks=true, colorlinks=true, linkcolor=magenta, citecolor=blue, urlcolor=blue, draft=false]{hyperref}
\usepackage[all]{hypcap}

\newcommand{\xabs}{{\tt xabs}\xspace}
\newcommand{\hot}{{\tt hot}\xspace}

\newcommand{\NH}{\ensuremath{N_{\mathrm{H}}}\xspace}

\newcommand{\xmm}{{\it XMM-Newton}\xspace}
\newcommand{\chandra}{{\it Chandra}\xspace}
\newcommand{\spitzer}{{\it Spitzer}\xspace}

\newcommand{\spex}{{\textsc{Spex}}\xspace}

\newcommand{\zi}{{\ensuremath{\zeta_{\rm{cryst}}}}\xspace}
\newcommand{\zii}{{\ensuremath{\zeta_{\rm{oliv}}}}\xspace}
\newcommand{\ziii}{{\ensuremath{\zeta_{\rm{Mg}}}}\xspace}
\newcommand{\daic}{{\ensuremath{{\rm \Delta AIC}}}\xspace}
\newcommand{\daicfour}{{\ensuremath{{\rm \Delta AIC < 4 }}}\xspace}
\newcommand{\daicten}{{\ensuremath{{\rm \Delta AIC < 10 }}}\xspace}

\newcommand{\sii}{\ion{Si}{i}\xspace}

\newcommand{\sixiii}{\ion{Si}{xiii}\xspace}
\newcommand{\mgxii}{\ion{Mg}{xii}\xspace}
\newcommand{\nex}{\ion{Ne}{x}\xspace}

\defcitealias{Rogantini19}{Paper I}
  
\definecolor{Gray}{gray}{0.95}
\definecolor{Cyan}{rgb}{0.9,1,1}
\definecolor{Red}{rgb}{0.992, 0.8, 0.69}

\makeatletter
\renewcommand*\aa@pageof{, page \thepage{} of \pageref*{LastPage}}
\makeatother

\begin{document}

   \title{Magnesium and silicon in interstellar dust: an X-ray overview}

   \subtitle{}
   \titlerunning{Magnesium and silicon in interstellar dust}
   \authorrunning{D. Rogantini, E. Costantini et al.}

   \author{D. Rogantini,
          \inst{1,2}
          E. Costantini,\inst{1,2}
          S.T. Zeegers,\inst{3}
          M. Mehdipour,\inst{1}
          I. Psaradaki,\inst{1,2}
          A.J.J. Raassen,\inst{1,2}
          C.P. de Vries,\inst{1}
           \and
          L.B.F.M. Waters\inst{1,2}
          }

   \institute{SRON Netherlands Institute for Space Research, Sorbonnelaan 2, 3584 CA Utrecht, the Netherlands\\
              \email{d.rogantini@sron.nl}
             \and 
             Anton Pannekoek Astronomical Institute, University of Amsterdam, P.O. Box 94249, 1090 GE Amsterdam, the Netherlands
             \and
             Academia Sinica Institute of Astronomy and Astrophysics, 11F of AS/NTU  Astronomy-Mathematics Building, No.1, Section 4, Roosevelt Rd, Taipei10617, Taiwan, ROC
             }

    \date{Date: 07/07/2020}

 
  \abstract
   {The dense Galactic environment is a large reservoir of interstellar dust. Therefore, this region represents a perfect laboratory to study the properties of the cosmic dust grains. X-rays are the most direct way to detect the interaction of light with dust present in these dense environments.}
   {The interaction between the radiation and the interstellar matter imprints specific absorption features in the X-ray spectrum. We study them with the aim of defining the chemical composition, the crystallinity and structure of the dust grains which populate the inner regions of the Galaxy.}
   {We investigate the magnesium and the silicon K-edges detected in the \chandra/HETG spectra of eight bright X-ray binaries, distributed in the neighbourhood of the Galactic centre. We model the two spectral features using accurate extinction cross sections of silicates, that we have measured at the synchrotron facility Soleil, France.}
   {Near the Galactic centre magnesium and silicon show abundances similar to the solar ones and they are highly depleted from the gas phase ($\delta_{\rm{Mg}}>0.90$ and $\delta_{\rm{Si}}>0.96$). We find that amorphous olivine with a composition of $\rm MgFeSiO_{4}$ is the most representative compound along all lines of sight according to our fits. The contribution of Mg-rich silicates and quartz is low (less than $10\%$). On average we observe a percentage of crystalline dust equal to 11\%. For the extragalactic source LMC X-1, we find a preference for forsterite, a magnesium-rich olivine. Along this line of sight we also observe an underabundance of silicon $A_{\rm Si}/A_{\rm LMC} = 0.5\pm0.2$.}
   {}

    \keywords{astrochemistry --
                X-rays: binaries --
                  X-rays: individuals: LMC X-1 --
                    X-rays: ISM --
                      dust, extinction 
               }

   \maketitle
%

\section{Introduction}

Over the last 20 years, X-ray absorption spectroscopy demonstrated to be a successful tool to study the chemical and physical properties of interstellar dust. Pioneering works \cite[e.g.][]{Lee02,Takei03,Costantini05} already detected dust signatures in the X-ray spectra taken with the high-resolution spectrometers aboard \xmm and \chandra, at that time newly launched (both in 1999). These dust features generically known as X-ray Absorption Fine Structures \citep[XAFS,][]{Newville04}, appear as a modulation of the region beyond the energy of photoelectric absorption edge. They are produced by the interaction between a photoelectron wave and all the other waves backscattered by neighbouring atoms in the solid lattice. XAFS are sensitive to the composition and structure of the absorber and, therefore, they represent a unique probe for investigating the chemistry, crystallinity and size distribution of the interstellar dust. \\ 

\noindent
Most of the common metals included in the cosmic dust show a photoelectric edge in the X-ray band $0.2-8$~keV. This allows us to investigate the nature of different dust species like silicates, carbonaceous material, sulphides and oxides. Nowadays, accurate extinction cross-sections of several interstellar dust analogues are available. It is possible to characterise the XAFS for multiple photoelectric K-shells: in particular the K edges of carbon \citep{Bilalbegovic18, Costantini19}, oxygen \citep[][Psaradaki et al., in prep.]{Costantini12}, magnesium \citep{Rogantini19}, silicon \citep{Zeegers17,Zeegers19}, iron \citep{Lee05,Rogantini18}, aluminium, sulfur and other low abundant elements \citep{Costantini19} plus the L-edges of iron \citep{Lee09,Westphal19}. The extinction cross-sections are calculated from the optical constants specific to the material. These are mainly measured using synchrotron radiation or through electron energy loss spectroscopy \citep[e.g. with the electron microscope,][]{Egerton09}. An alternative is to use the density functional theory \citep[DFT,][]{Jones15}, a computational quantum mechanical modelling which allows the prediction and calculation of material behaviour based on the relative electron density.\\

\noindent
The spectra of bright low-mass X-ray binaries, lying in the plane of our Galaxy, are perfect laboratories to investigate the dust fine structures. Indeed, relative large column densities guarantee an optimal optical depth of the edge and high fluxes are necessary to ensure a sufficient signal-to-noise ratio to distinguish the dust features. \cite{Zeegers19} analysed the line of sight of 9 bright X-ray sources nearby the Galactic centre. They found that most of the spectra can be well fit by amorphous olivine ($\rm MgFeSiO_{4}$).
Nonetheless, interstellar silicates are expected to exist in various forms in the interstellar medium. For example, \cite{Kemper04}, \cite{Chiar06}, and \cite{Min07} compared laboratory spectra with mid-infrared observations and found, in fact, that a mixture of olivine and pyroxene dust models fit the $\rm \sim{9}\, \mu m$ feature.\\

\noindent
The broad, smooth and featureless infrared bands around $\rm  9.7\ and\ 18\ \mu m$ also suggest that most of the interstellar silicates are amorphous \citep{Li01a,Li07,Molster10}. From a direct comparison of the Sgr A$^{*}$ spectrum with theoretical spectra for pure silicates, \cite{Kemper04} found that only less than $2.2\%$ of the dust should have a crystalline order. \cite{Li07} concluded that the allowed degree of crystallinity would be $\sim 5\%$ considering the effect of the ice mantle coating the silicates cores on the determination of the crystallinity degree of silicates. Recent X-ray observations found a higher amount of crystalline dust, between 3-20\%, in the region near the centre of the Galaxy \citep{Rogantini19,Zeegers19}. Moreover,  crystalline dust has been observed in a variety of environments, from proplanetary disks \citep{Honda03,Natta07} to diffuse interstellar medium \citep{Westphal14}. The precise amount of crystalline dust and its survival in the interstellar medium are still widely debated in the scientific community. \\

\noindent
X-ray absorption spectroscopy provides the possibility to study the composition of the dust grains in different environments \citep{Lee02,Draine03,Ueda05,Valencic09,Costantini12,Pinto13,Corrales16,Hoffman16,Schulz16,Zeegers19}. Dust grains undergo a cyclic process of production, growth and destruction which may change their properties. Dust particles mainly condense in the vicinity of late-type stars, in nova and supernova ejecta. In harsh and turbulent regions, the grains are exposed to radiation which may reprocess and destroy them. Only refractory cores would survive in these environments \citep{Whittet02}. Differently, in the dark and cold molecular clouds of the Galaxy, grains are shielded from the radiation. Consequently, we expect to observe larger grains in these environments with a more complex structure. In these low temperature environments ices may accrete onto these pre-existing refractory cores. The ice mantle is the chemical laboratory for the production of more elaborate molecules.\\
 Due to their high penetrating power, X-rays can be used to probe the dust grain properties in Galactic regions characterised by different densities.\\

\noindent
From studying the metallicity of the B-type stars, Cepheids and open clusters, the abundances of the common elements show, in general, a gradient as a function of the distance from the Galactic centre \citep{Rolleston00,Pedicelli09,Genovali14}. These variations would be a consequence of the successive generations of stars having enriched the interstellar matter. Thus, abundance investigation can provide a key to understanding the formation and chemical evolution of galaxies. However, only a few measurements are available within 5 kpc from the Galactic centre \citep{Rich17,Schultheis19}. With X-rays, we can investigate directly the abundances and depletions of several metals in the inner part of the Galaxy and therefore extend the characterisation of the metallicity gradient observed in the Galactic plane.\\

\noindent
In this study, we focus on the magnesium and silicon K-edges (located at 1.3 and 1.84 keV, respectively) through which we study the denser region in the central part of the Galaxy. Currently, the \chandra High-Energy Transmission Grating Spectrometer (HETGS) represents the best instrument for studying simultaneously the two absorption K-edges thanks to the high effective area and energy resolution of its gratings, MEG and HEG, in the $1-2\,\rm{keV}$ energy range. \\
Magnesium and silicon are two abundant\footnote{The abundances are often given in logarithmic scale relative to $\NH = 10^{12}$ For element X, $$\log A_{\rm X} = 12 + \log (N_{\rm{X}}/\NH)\ .$$ It is often convenient to express the abundances of the common metals on a linear scale relative to $\NH=10^6$, i.e. in parts per million (ppm) relative to hydrogen: $A_{\rm{Mg}}= 39$ ppm and $A_{\rm{Si}}=35$ ppm.} metals in the interstellar matter: respectively, they are the eight and the ninth most abundant element in the Universe with $\log A_{\rm{Mg}} = 7.599 $ and $\log A_{\rm{Si}} = 7.586$ \citep{Lodders10}. Magnesium is mostly the product of carbon and neon burning in core-collapse supernovae, whereas silicon is produced by oxygen shell burning \citep{Arnett85,Thielemann85}. In the interstellar matter, they are both highly depleted from the gas phase: in dense environments, more than $90\%$ of their mass is thought to have been locked in dust particles \citep{Jenkins09,Palme14,Zhukovska18}. The depletion\footnote{We define depletion as the ratio of the dust abundance to the total abundance of a given element, i.e. both gas and dust.} of magnesium and silicon is adequately explained by the formation of a mixture of both iron-rich and iron-poor silicates \citep{Jones87,Jones00,Kimura03,Mattsson19}. Other silicon- and magnesium-bearing species, such as silicon carbide (SiC), spinel ($\rm MgAl_2O_4$), gehlenite ($\rm Al_2Ca_2SiO_7$) and diopside ($\rm MgCa(SiO_3)_2$) are too rare to make a significant contribution \citep{Jones07}. Thus, magnesium and silicon share the same depletion trend as a function of the environment. \\
\noindent
In our pilot work we showed the analysis of Mg and Si K-edges of the bright X-ray binary GX 3+1 \citep[][henceforth, \citetalias{Rogantini19}]{Rogantini19}. Here, we expand the number of sources to characterise the cosmic dust properties and quantify the abundances and depletion of silicon and magnesium in the dense neighbourhood of the Galactic centre. The source sample is presented in Section \ref{sec:sample_analysis}, together with the data reduction and analysis of the \chandra observations. The simultaneous fit of the Si and Mg absorption K-edges with both gas and dust models is described in Section \ref{sec:mg_si_edges}. We discuss the results and the properties of the interstellar dust in Section \ref{sec:discussion}. Finally, in Section \ref{sec:conclusion}, we provide our conclusions and a summary of our results. Throughout the paper, for the fitting process we use the $C$-statistics \citep{Cash79,Kaastra17} unless otherwise stated. The errors quoted are for the $68\%$ confidence level. In the analysis, we adopt the protosolar abundances tabulated by \cite{Lodders10}.

\section{Sample and analysis}
\label{sec:sample_analysis}

\subsection{Source sample selection} 
\label{sec:source_selection}
We select our sources from the \textit{Chandra} Data Archive\footnote{See \url{http://cxc.harvard.edu/cda/}} following a criterion to optimise the detections of the two edges. The primary selection is based on the hydrogen column density ($N_{\rm H}$). The column density towards the source is closely related to the optical depths of the edges. The range $0.5-5 \times10^{22}\,\rm cm^{-2}$ allows an adequate level of X-ray transmittance for our analysis. Second, we select the sources with a flux larger than $1\times10^{-11}\rm\,erg\,cm^{-2}\,cm^{-1}$ in the soft X-ray energy band $0.5-2\,\rm keV$ and enough exposure time, in order to ensure a high signal-to-noise ratio to observe the fine structures of the edges.\\
In this study, we use bright lowmass X-ray binaries (LMXBs). They obey the two conditions above and their spectrum, generally, does not host emission lines in the soft energy band. Moreover, we verify that the sources in our sample are persistently bright, in order to make the best use of the satellite exposure times.\\
Finally, we select only the observations taken in timed exposure (TE) mode. The continuous clocking (CC) mode is not suitable for analysing the magnesium and silicon K-edge. Absorbed spectra taken in CC-mode are affected by the contribution of the two-dimensional scattering halo around the source collapse in a one-dimensional image. Its contribution is hard to disentangle from the dispersed spectrum, particularly in the absorption edge regions (we refer to the \textit{Chandra} Proposers' Observatory Guide\footnote{\url{http://cxc.harvard.edu/proposer/POG/pdf/MPOG.pdf}}, version 21.0).
In total, we select seven Galactic X-ray binaries plus the brightest X-ray source in the Large Magellanic Cloud, LMC X-1. These sources are summarised in Table \ref{tab:sample} where we indicate the obsID, exposure time, and average count rate for each observation, plus the Galactic coordinates and distance of the sources, as reported in the literature. 

%
\begin{table}
\caption{X-ray binaries}             
\label{tab:sample}
\tiny
\centering
\setlength\tabcolsep{4.95pt}
\begin{tabular}{r c c c r r c }     
\hline\hline       
 & && & \multicolumn{2}{c}{Galactic coordinates}  &    \\ \cline{5-6}
       
\multirow{-2}{*}{Obsid} & \multirow{-2}{*}{Date}      &  \multirow{-2}{*}{Exp.}          &  \multirow{-2}{*}{Rate}        & \multicolumn{1}{c}{$l$} & \multicolumn{1}{c}{$b$} & \multirow{-2}{*}{Distance} \\

       &  UT  &   ks     & $c/s$  & \multicolumn{1}{c}{deg} & \multicolumn{1}{c}{deg} &  kpc  \\
\hline
\rowcolor{Gray}
\multicolumn{7}{c}{\object{GRS 1758-258}}        \\
\hline
\noalign{\vskip 1.5mm}              
2429  & 2001-03-24 & 29.6 & 17.2  & \multirow{2}{*}{$4.508$} & \multirow{2}{*}{$-1.361$} & \multirow{2}{*}{$8^{(a)}$}  \\
2750  & 2002-03-18 & 27.5 & 30.3 \\
\noalign{\vskip 1.5mm}
\hline
\rowcolor{Gray}
\multicolumn{7}{c}{\object{GRS 1915+105}}        \\
\hline
\noalign{\vskip 1.5mm}
660   & 2000-04-24 & 30.6 & 127.2 & \multirow{2}{*}{$45.366$} & \multirow{2}{*}{$+0.219$} & \multirow{2}{*}{$8.6\pm2.0^{(b)}$}  \\
7485  & 2007-08-14 & 48.6 & 150.3 \\
\noalign{\vskip 1.5mm}
\hline
\rowcolor{Gray}
\multicolumn{7}{c}{\object{GX 3+1}}              \\
\hline
\noalign{\vskip 1.5mm}
16492 & 2014-08-17 & 43.6 & 100.4 & \multirow{7}{*}{$2.294$} & \multirow{7}{*}{$+0.794$} & \multirow{7}{*}{$6.1^{(c)}$}  \\
16307 & 2014-08-22 & 43.6 & 102.2\\
18615 & 2016-10-27 & 12.2 & 67.6 \\
19890 & 2017-05-23 & 29.1 & 86.2 \\
19907 & 2016-11-02 & 26.0 & 70.0 \\
19957 & 2017-04-30 & 29.1 & 93.1 \\
19958 & 2017-05-21 & 29.1 & 86.4 \\
\noalign{\vskip 1.5mm}
\hline
\rowcolor{Gray}
\multicolumn{7}{c}{\object{GX 9+1}}             \\
\hline
\noalign{\vskip 1.5mm}
717   & 2000-07-18 & 9.0  & 165.2 & $9.077$ & $+1.154$ & $5^{(d)}$ \\
\noalign{\vskip 1.5mm}
\hline
\rowcolor{Gray}
\multicolumn{7}{c}{\object{GX 17+2}}         \\
\hline
\noalign{\vskip 1.5mm}
11088 & 2010-07-25 & 29.1 & 176.5 & $16.432$ & $+1.277$ & $9.1\pm0.5^{(e)}$ \\
\noalign{\vskip 1.5mm}
\hline
\rowcolor{Gray}
\multicolumn{7}{c}{\object{H 1742-322}}         \\
\hline
\noalign{\vskip 1.5mm}
16738 & 2015-06-11 & 9.2  & 16.8  & \multirow{4}{*}{$357.255$} & \multirow{4}{*}{$-1.833$} & \multirow{4}{*}{$8.5^{(f)}$}  \\
17679 & 2015-06-12 & 9.2  & 17.4 \\
17680 & 2015-06-13 & 9.2  & 18.0 \\
16739 & 2015-07-03 & 26.8 & 10.7 \\
\noalign{\vskip 1.5mm}
\hline
\rowcolor{Gray}
\multicolumn{7}{c}{\object{IGR J17091-3624}}     \\
\hline
\noalign{\vskip 1.5mm}
12406 & 2011-10-06 & 27.3 & 46.8  & \multirow{3}{*}{$349.525$} & \multirow{3}{*}{$+2.213$} & \multirow{3}{*}{$12^{(g)}$} \\
17787 & 2016-03-30 & 39.5 & 29.9  \\
17788 & 2016-04-30 & 38.8 & 28.5  \\
\noalign{\vskip 1.5mm}
\hline
\rowcolor{Gray}
\multicolumn{7}{c}{\object{LMC X-1}}             \\
\hline
\noalign{\vskip 1.5mm}
93    & 2000-01-16 & 18.9 & 26.5  & \multirow{11}{*}{$280.203$} & \multirow{11}{*}{$-31.516$} & \multirow{11}{*}{$48^{(h)}$} \\
11074 & 2010-01-02 & 17.2 & 24.2   \\
11986 & 2010-01-07 & 8.2  & 25.5   \\
11987 & 2010-01-18 & 18.6 & 22.3   \\
12068 & 2010-01-04 & 13.1 & 23.5   \\
12071 & 2010-01-09 & 4.3  & 25.7   \\
12069 & 2010-01-08 & 18.1 & 26.0   \\
12070 & 2010-01-10 & 17.4 & 25.7   \\
12089 & 2010-01-21 & 14.7 & 23.3   \\
12090 & 2010-02-26 & 14.0 & 23.4   \\
12072 & 2010-01-05 & 18.1 & 23.9   \\
\noalign{\vskip 1.5mm}
\hline                  
\end{tabular}
\tablebib{(a)~\citet{Soria11}: assumed distance; (b) \citet{Reid14}; (c) \citet{denHartog03}; (d) \citet{Iaria05}; (e) \citet{Galloway08}; (f) \citet{Steiner12}; (g) \citet{Court17}: assumed distance; (h) \citet{Orosz09}.}
\end{table}

\subsection{Data reduction}
\label{sec:data_reduction}
We obtain the observations for our analysis from the \chandra Transmission Gratings Catalog and Archive \citep[\textsc{TGCat}\footnote{See \url{http://tgcat.mit.edu/}},][]{Huenemoerder11}. For each observation, we select both HEG and MEG spectra and we combine separately the positive and negative first orders using the tool \texttt{combine\_grating\_spectra} included in the Chandra's data analysis system CIAO \citep[version 4.11,][]{Fruscione06}. For the brightest sources, the grating spectra taken in TE mode are affected by photon pile-up. The bulk of pileup events comes from the MEG 1st orders and affects, in particular, the harder part of the spectra. Both magnesium and silicon edge regions are relatively less affected, as in these regions the spectrum is depressed due to the interstellar medium column density (see also Appendix \ref{app:broadband}).\\
All spectra show an apparent 1-resolution-element-bin-wide excess at $6.741$ \AA. This excess has been previously observed also in other sources, with different interpretations: an emission line from \ion{Si}{XIII} \citep[e.g.][]{Iaria05}, dust scattering peak \citep[][and \citetalias{Rogantini19}]{Schulz16} or instrumental effect \citep{Miller02,Miller05}. Here, we adopt the latter interpretation (justified in Appendix \ref{app:6.74}). Therefore, we add to the continuum a delta line model centred at $6.741\pm0.001$ {\AA}. This value has been estimated using the spectrum of three bright and well-known low mass X-ray binaries, namely GX 9+9, Cyg X-2, and 4U~1820-30, plus the blazar Mrk 421 (see Appendix \ref{app:6.74}). 

\subsection{Continuum and absorption}
\label{sec:continuum}
In order to study the dust absorption, first we characterise the broadband spectrum and then we focus on the fit of the magnesium and silicon K-edges with our dust extinction models. \\
To fit the underlying continuum of each source we use the spectral analysis code \spex\footnote{\url{10.5281/zenodo.1924563}} version 3.05 \citep{Kaastra96,Kaastra18}. We use a two-component spectral model consisting of a thermal component in the soft end and non-thermal component in the hard end of the spectral bandpass. In order to obtain the best fit of the continuum we test different combinations of several \spex emission models: blackbody \citep[\texttt{bb},][]{Kirchhoff60}, disk blackbody \citep[\texttt{dbb},][]{Shakura-Sunyaev73,Mitsuda84}, modified blackbody \citep[\texttt{mbb},][]{Rybicki86,Kaastra89} as thermal models and power-law (\texttt{pow}) and Comptonisation \citep[\texttt{comt},][]{Titarchuk94} as non-thermal models.\\
The absorption by cold gas is given by the multiplicative model \texttt{hot} \citep{dePlaa04,Steenbrugge05} fixing the electron temperature at the lower limit, that is $kT_{\rm e}=0.5\,\rm{eV}$. By default, \spex adopts protosolar abundances for the gas phase \citep{Lodders10}. For LMC X-1, we apply the typical element abundances found in the Large Magellanic Cloud, listed with the relative references in Table \ref{tab:lmc_abundance}.\\
In this study we update the neutral magnesium and silicon cross-sections with respect to the official release of \spex, adding the resonance transitions, $1s\rightarrow np$, calculated using both the Flexible Atomic Code \citep{Gu08} and COWAN code \citep{Cowan95}, respectively. We present them in Appendix \ref{app:neutral_silicon}.\\
For sources with multiple data sets, we fit simultaneously the continuum of the different observations by coupling the absorption by neutral gas in the interstellar matter that we assume to be constant. In Table \ref{tab:result_fit} we summarise the parameter values of best model for each source. \\
Also, we test whether there is ionised gas along the line of sight towards the sources. In particular, we investigate the presence of collisionally ionised gas, by adding to the model an extra \texttt{hot} component, and photo-ionised gas, by applying the \texttt{xabs} model \citep{Steenbrugge03} to the continuum. For an accurate modelling, it is essential to distinguish absorption lines due to ionised gas, especially if they appear near the edges, where they can be confused with absorption features by neutral gas or cosmic dust. We find the presence of photo-ionised gas in outflow in GRS 1915+105 with an outflow velocity of $v \sim 145 \pm 9\,\rm km / s$ and logarithmic ionisation parameter\footnote{Here, the ionisation index is defined as $\xi = L/(nr^{2})$ where $L$ is the luminosity of the source, $n$ the density of the gas and $r$ the distance between the ionising source and the absorbing gas.} $\log \xi = 3.72 \pm 0.02$ and $N_{\rm{H}} = (5.9\pm0.7)\times 10^{20}\ \rm cm^{-2}$. These values are consistent, within the uncertainties, with the results obtained by \cite{Ueda09}.\\ Along the line of the sight of the remaining sources, we do not find significant evidence of either photo-ionised and collisional-ionised gas.\\

\section{The magnesium and silicon edge models}
\label{sec:mg_si_edges}

\begin{figure*}
   \centering
     \includegraphics[width=.98\hsize]{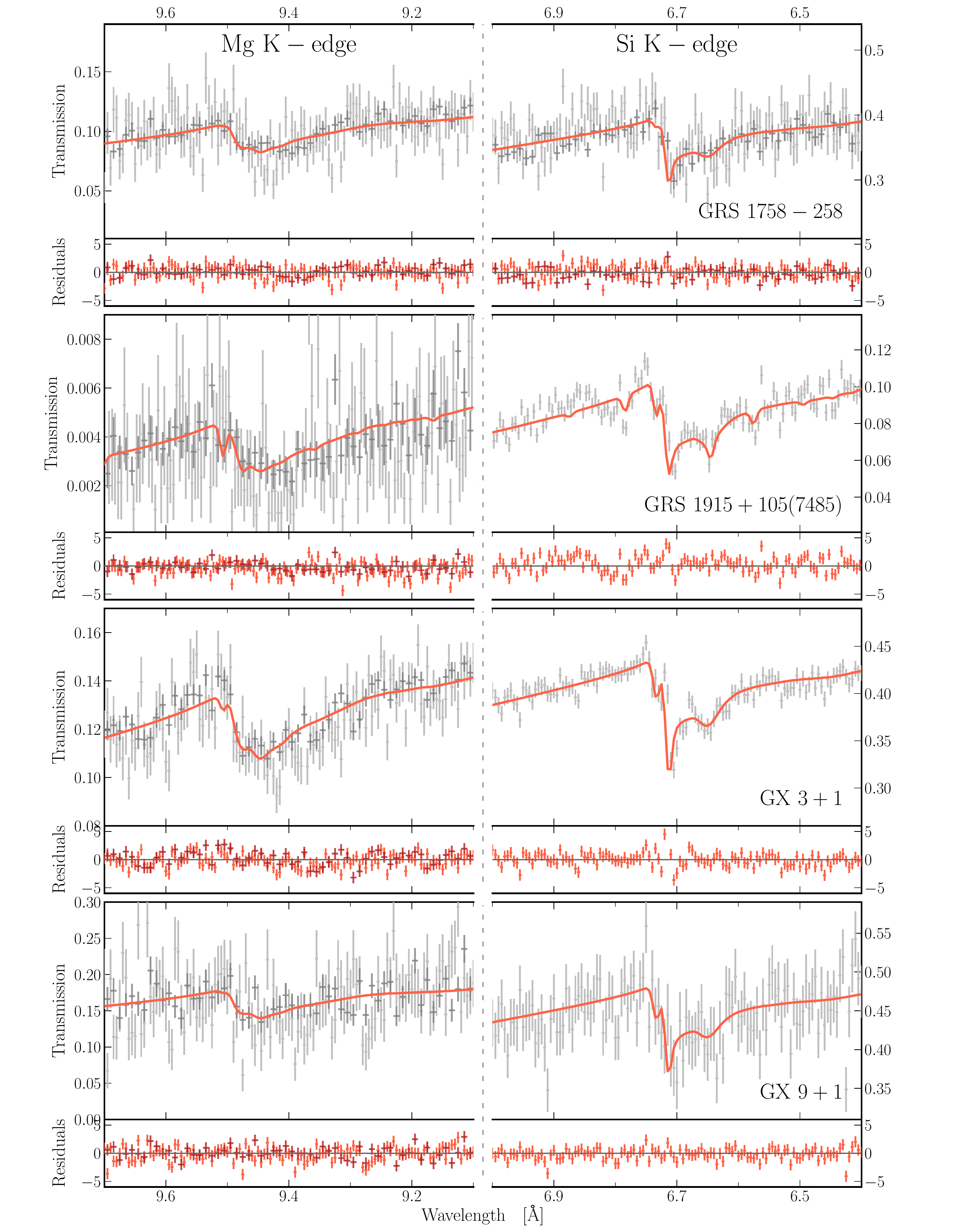}\\
      \caption{Zoom-in on the magnesium (on the left column) and silicon K-edge (on the right column). The HEG and MEG data are respectively shown in light and dark grey. The solid red line represents the best fit whose dust composition is specified in Table \ref{tab:dust_properties}. In bottom panels we show the residuals defined as $\rm (data-model)/error$. The data are stacked and binned for display purpose. For GRS 1915+105 we display only observation ID 7485, which shows also photoionisation lines (see Appendix \ref{app:broadband} for further details).}
         \label{fig:edges1}
   \end{figure*}

\begin{figure*}
   \centering
     \includegraphics[width=.98\hsize]{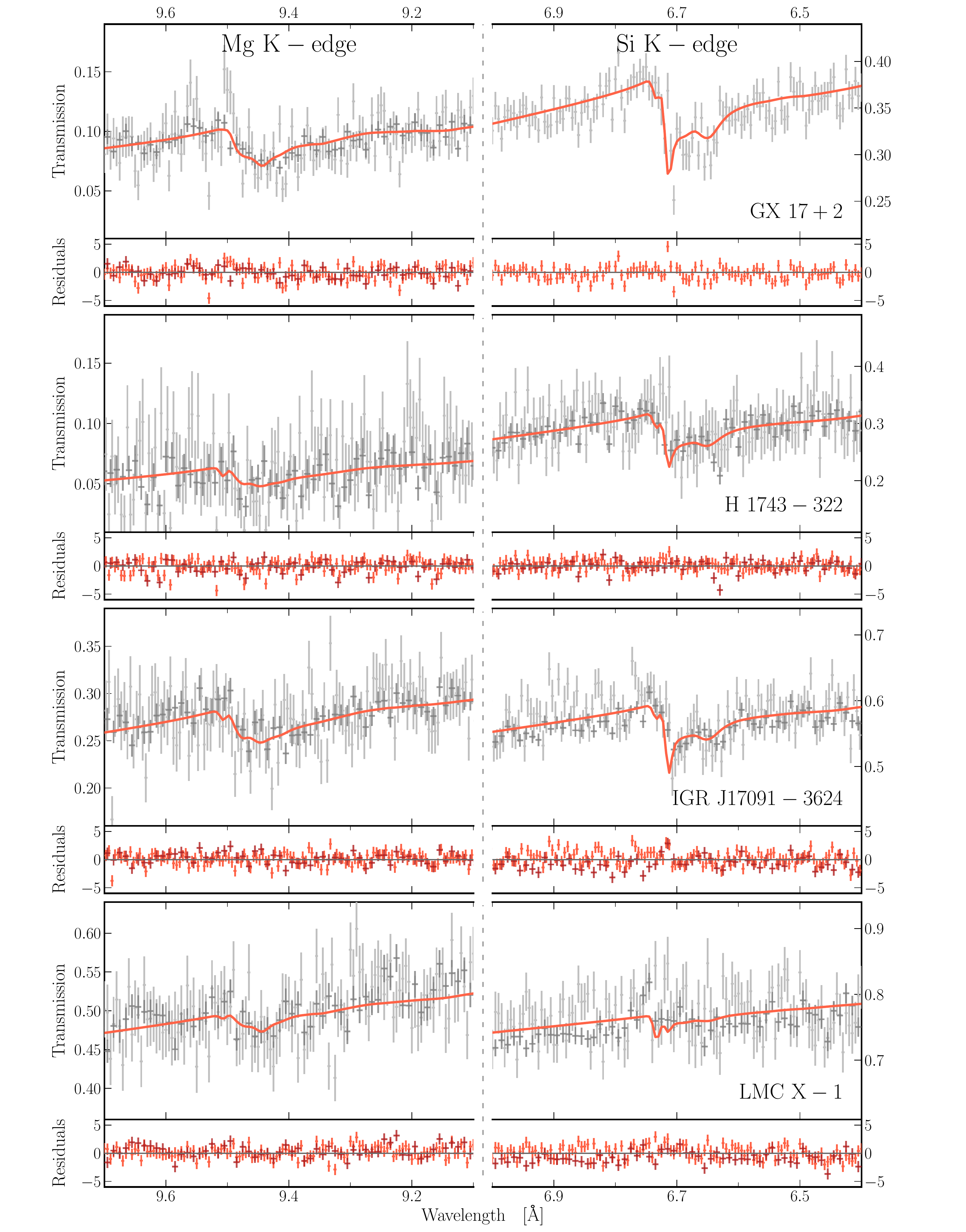}
      \caption{Same as Figure \ref{fig:edges1}.}
         \label{fig:edges2}
   \end{figure*}

\begin{figure*}
   \centering
     \includegraphics[width=.85\textwidth]{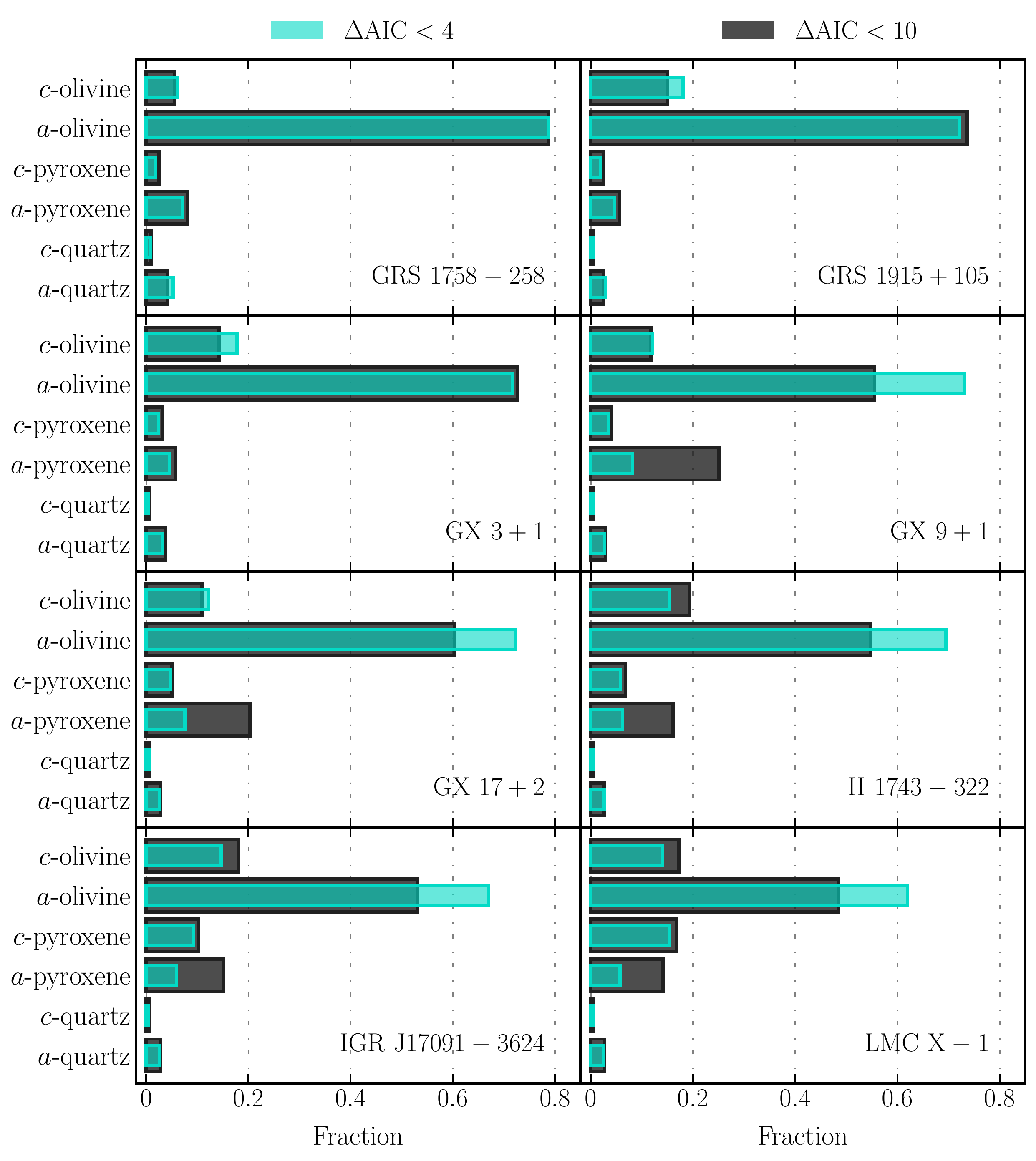}
      \caption{Relative fraction for the different dust species calculated considering models with \daicfour in light blue and \daicten in dark gray. We group the compounds by their structure and crystallinity (see Section \ref{sec:mg_si_edges} for details). We use the abbreviations $c-$ for crystalline and $a-$ for amorphous.}
         \label{fig:aic_plot}
   \end{figure*}

In \citetalias{Rogantini19}, we show how the simultaneous fit of multiple edges of different elements allows us to better constrain, with respect to a single-edge modelling, the chemical properties and size of the interstellar grains, limiting the possible degeneracies of the fit. These edges are the result of the absorption by cold gas together with the interstellar dust present along a relatively dense line of sight. In the case of magnesium and silicon, we expect a large contribution by cosmic dust since a large fraction of these elements is included in dust grains. Whereas gaseous contribution is already modelled by the \texttt{hot} component, it is necessary to add to the broadband model the \texttt{AMOL} model \citep{Pinto10} for shaping the cosmic dust scattering and absorption. In order to evaluate the dust-to-gas ratio, we set these two models free to compete for the fitting of the edges. \\
In our analysis, we use dust extinction models based on accurate laboratory measurements. These models include both the scattering and absorption cross-sections and we summarise them in Table \ref{tab:compounds}, where we specify their chemical formula and crystallinity. The laboratory measurements and post-processing are explained in \cite{Zeegers17,Zeegers19} and \citetalias{Rogantini19}. Here, we assume the Mathis-Rumpl-Nordsieck \citep[MRN,][]{Mathis77} dust grain size distribution, which follows a power-law distribution, $dn/da\propto a^{-3.5}$ where $a$ is the grain size, with minimum and maximum cut-offs of 0.005 and $0.25~\mu\rm m$, respectively. \\
The \texttt{AMOL} model allows for four dust compounds to be tested in a given fitting run. Thus, we test all the possible combinations of the 14 dust models following the method of \cite{Costantini12} and obtaining 1001 different models to fit for each source.\\ 
We select as best fit the dust mixture which presents the minimum $C$-statistic value among all the models. 
In Figures \ref{fig:edges1} and \ref{fig:edges2} we show, for all the sources, the $Chandra$ data and their best fit around the magnesium and silicon edges and residuals to the best fit. The dust mixtures which characterise the best fits are listed in Table \ref{tab:dust_properties} with the relative column densities. Following the procedure presented in \citetalias{Rogantini19}, we select the models statistically similar to the best fit through the \emph{Aikake Information Criterion} \citep[AIC,][]{Akaike74,Akaike98}. In particular, the $\Delta \rm AIC_{i}$, namely the difference between the $i$-model and the selected best model, allows a meaningful comparison and ranking of the candidate's models. \\
Based on the criteria presented in \cite{Burnham02} we consider the $i$-models with $\rm AIC_{i} < 4$ statistically comparable to the selected best model. Instead, models with a $\daic > 10$ can be ommited from further consideration. The overall analysis of the selected models permits to understand and define the characteristics of the most representative dust compounds and at the same time to rule out the dust species which fail in describing the magnesium and silicon edges. \\
In Figure \ref{fig:aic_plot}, we show the relative fractions of the dust compounds. For clarity, we cluster the compounds with similar crystallinity and structure. In particular, compounds number (1, 2, 4) are grouped as crystalline olivine ($c$-olivine); compound number 3 as amorphous olivine ($a$-olivine); (6, 8, 11) as $c$-pyroxene; (5, 7, 9, 10) as $a$-pyroxene; 12 as $c$-quartz; (13, 14) as $a$-quartz. In the bar chart, we show the models with \daicfour in light blue together with the models with \daicten, in black, which include fits with has a less significance compared to the best fit.\\


%
\begin{table}
\caption{List of interstellar dust extinction models. 
        }             
\label{tab:compounds}      
\centering
\renewcommand{\arraystretch}{1.1}
\begin{tabular}{r l l c}        
\hline\hline           
\# & Name & Chemical formula & Form \\    
\hline                        
\noalign{\vskip 1.0mm}
 1  &  Forsterite       & Mg$_{2}$SiO$_4$                         & crystalline \\
 2  &  Fayalite         & Fe$_{2}$SiO$_4$                         & crystalline \\
 3  &  Olivine          & MgFeSiO$_4$                             & amorphous   \\
 4  &  Olivine          & Mg$_{1.56}$Fe$_{0.4}$Si$_{0.91}$O$_4$   & crystalline \\
 5  &  Enstatite        & MgSiO$_3$                               & amorphous   \\
 6  &  Enstatite        & MgSiO$_3$                               & crystalline \\
 7  &  En60Fs40         & Mg$_{0.6}$Fe$_{0.4}$SiO$_3$             & amorphous   \\
 8  &  En60Fs40         & Mg$_{0.6}$Fe$_{0.4}$SiO$_3$             & crystalline \\
 9  &  En75Fs25         & Mg$_{0.75}$Fe$_{0.25}$SiO$_3$           & amorphous   \\
 10 &  En90Fs10         & Mg$_{0.9}$Fe$_{0.1}$SiO$_3$             & amorphous   \\
 11 &  En90Fs10         & Mg$_{0.9}$Fe$_{0.1}$SiO$_3$             & crystalline \\
 12 &  Quartz           & SiO$_2$                                 & crystalline \\
 13 &  Quartz           & SiO$_2$                                 & amorphous   \\
 14 &  Quartz           & SiO$_2$                                 & amorphous   \\
\noalign{\vskip 1.0mm}
\hline                                   
\end{tabular}
\tablefoot{The nomenclature En(x)Fs(1-x) indicates the fraction of magnesium (or iron) included in the compound. "En" stands for enstatite (the Mg-pure pyroxene, $\rm{Mg}\rm{Si}\rm{O}_{3}$) and "Fs" for ferrosilite (the Fe-pure pyroxene, $\rm{Fe}\rm{Si}\rm{O}_{3}$). The two amorphous quartz present two different level of glassiness: compound number 13 has an intermediate amorphous form whereas compound 14 has a full amorphous structure.}
\end{table}

%
%
\begin{table*}
\caption{Best fitting dust compounds for each source and their derived properties.}            
\label{tab:dust_properties}
\footnotesize
\centering
\setlength\tabcolsep{4.95pt}        
\begin{tabular}{l c c c c c c c c c c c c }   
\hline\hline       
Dust compound & GRS1758-258 & GRS1915+105 & GX 3+1& GX 9+1 & GX 17+2 & H1743-322 & IGRJ17091-3624 & LMC X-1\\
\noalign{\vskip 1.5mm}
& \multicolumn{8}{c}{ Dust column densities [$10^{17}\,\rm cm^{-2}$]} \\
\hline                    
{\scriptsize(1)} $c$-Forsterite        &  --          &   --       &   --         &   --    & $ 0.6\pm0.3$ &   --     &   --          & $ 0.45\pm0.08$     \\
{\scriptsize(2)} $c$-Fayalite          &  --          & $ 6\pm1$   &   --         & $ <0.3$ & $ <0.8$      & $ 2\pm1$ &   --          &   --               \\
{\scriptsize(3)} $a$-Olivine           & $7.7\pm0.4$  & $ 15\pm1 $ & $ 7.0\pm0.1$ & $ 6\pm2$& $ 7.8\pm0.7$ & $ 8\pm2$ & $ 4.1\pm0.1$  & $ 0.33\pm0.09$     \\
{\scriptsize(4)} $c$-Olivine           &  --          &   --       &   --         &   --    &   --         &   --     &   --          & $ <0.23$           \\
{\scriptsize(5)} $a$-Enstatite         &  --          &   --       &   --         & $ <2.9$ &   --         &   --     &   --          &   --               \\
{\scriptsize(6)} $c$-Enstatite         &  --          &   --       &   --         &   --    &   --         &   --     &   --          & $ <0.05$           \\
{\scriptsize(7)} $a$-En60Fs40          &  --          &   --       &   --         &   --    &   --         &   --     &   --          &   --               \\
{\scriptsize(8)} $c$-En60Fs40          &  --          &   --       &   --         &   --    &   --         &   --     &   --          &   --               \\
{\scriptsize(9)} $a$-En75Fs25          & $ <0.9 $     & $  <0.2 $  &   --         &   --    & $ <0.6$      &   --     & $ <0.4$       &   --               \\
{\scriptsize(10)} $a$-En90Fs10         &  --          &   --       &   --         &   --    &   --         &   --     &   --          &   --               \\
{\scriptsize(11)} $c$-En90Fs10         &  --          &   --       &   --         &   --    &   --         &   --     &   --          &   --               \\
{\scriptsize(12)} $c$-Quartz           & $0.6\pm0.4$  &   --       & $ <0.08$     & $ <0.4$ &   --         & $ <0.7$  & $ <0.09$      &   --               \\ 
{\scriptsize(13)} $a$-Quartz           &  --          &   --       & $ 1.4\pm0.2$ &   --    &   --         & $ <0.1$  & $ 0.4\pm0.3$  &   --               \\ 
{\scriptsize(14)} $a$-Quartz           & $<0.3$       & $ <0.1$    & $ <0.06$     &   --    &   --         &   --     &   --          &   --               \\ 
\hline
\noalign{\vskip 0.75mm}
\multicolumn{1}{c}{Total}              & $ 8^{+2}_{-1}$ & $ 21\pm2$ & $ 8.4\pm0.3$ & $ 6^{+5}_{-2}$ & $8^{+2}_{-1}$ & $ 10\pm3$ & $ 4.5^{+0.8}_{-0.4}$ & $ 0.8^{+0.4}_{-0.2}$ \\
\noalign{\vskip 0.75mm}
\hline  
\multicolumn{1}{c}{$ \zeta_{1} $}      & $ 0.07\pm0.04 $ & $ 0.27\pm0.05 $ & $ <0.01       $ & $ <0.07       $ & $ 0.11\pm0.07 $ & $ 0.21\pm0.18 $ & $ <0.02       $ & $ >0.73 $ \\
\multicolumn{1}{c}{$ \zeta_{2} $}      & $ >0.94       $ & $ >0.98       $ & $ 1           $ & $ >69         $ & $ 0.96\pm0.03 $ & $ 1      $ & $ 0.95\pm0.04      $ & $ >0.92 $ \\
\multicolumn{1}{c}{$ \zeta_{3} $}      & $ 0.50\pm0.03 $ & $ 0.36\pm0.02 $ & $ 0.50\pm0.01 $ & $ 0.58\pm0.14 $ & $ 0.52\pm0.08 $ & $ 0.39\pm0.08 $ & $ 0.51\pm0.02 $ & $ >0.76 $ \\
\hline 

\end{tabular}
\tablefoot{In the upper part of the table we report the column density for each dust species (see Table \ref{tab:compounds}) and we calculate the total dust column density. In the bottom part we show the value of \zi, \zii, and \ziii as defined in the text. Errors given on parameters are $1\sigma$ errors.}
\end{table*}

\section{Discussion}
\label{sec:discussion}

\subsection{Mineralogy of the dust toward the Galactic center}
The properties of the cosmic dust derived through the fits of the eight X-ray binaries are listed in Table \ref{tab:dust_properties}. The analysis of the sample shows a clear preference for amorphous olivine (compound number 4) in the best fit and all the fits with $\daic<4$. For every source, amorphous olivine represents more than $60\%$ of the total amount of the dust reaching $98\%$ of the total for GRS 1758-258. In the bottom part of Table \ref{tab:dust_properties} we report the quantities \zi, \zii, and \ziii, which, respectively, give information about the crystallinity, structure and chemistry of the cosmic dust. Following the notation used by \cite{Zeegers19}, they are defined as:
\begin{align}
\label{eq:zetas}
\begin{split}
\zi     &= \rm crystalline\ dust  / (crystalline\ dust + amorphous\ dust)\ , \\
\zii    &= \rm olivine            / (olivine + pyroxene)\ ,                  \\
\ziii   &= \rm Mg_{dust}          / (Mg_{dust} + Fe_{dust})\ ,
\end{split}
\end{align}

\noindent
where with olivine we indicate the sum of compounds number 1-4 of Table \ref{tab:compounds}, whereas pyroxene represents compounds 5-11. $\rm Mg_{dust}$ and $\rm Fe_{dust}$ indicate, respectively, the magnesium and iron content of interstellar dust.\\
The quantities \zi, \zii, \ziii together with the chemical composition represent the dust properties accessible by X-ray absorption spectroscopy and we discuss them separately below. The extragalactic source LMC X-1 will be discussed in Section \ref{sec:lmcx-1}. 

\subsubsection{Crystallinity}
\label{sec:crystallinity}

In our analysis the best fits show a crystallinity \zi between $0-0.3$. Selecting the fits of X-rays sources with higher quality data, GX 3+1, GX 9+1, GX 17+2, and considering the uncertainties on the values, we observe crystallinity below 0.2. In particular for GX 3+1 we obtain a crystallinity upper limit of 0.01. The other sources in the analysis, show similar results, although for some of them, the \zi value is affected by large uncertainties because of the quality of the data. The microquasars GRS 1915+105 is a possible outlier, showing the largest amount of crystalline dust with $\zi = 0.27\pm0.05$. \\
Even considering the uncertainties on the parameters \zi, the average crystallinity fraction, $\langle \zi \rangle \sim 0.11$, is larger than the ones observed at longer wavelengths, in particular in the infrared. The smooth shape of the $\sim 9\ \mu\rm{m}$ and $\sim 18\ \mu\rm{m}$ absorption features suggests that less than $2.2\%$ of the dust in the interstellar medium appears to be crystalline \citep{Kemper04}. Our model set contains both the crystalline and amorphous counterpart of all the compounds except for two of them (see Table \ref{tab:compounds}). This may introduce some bias on our estimation of the amount of crystallinity along the line of sight of the X-ray binaries. In particular, this seems the case for GRS 1915+105 where the crystalline percentage is driven by the fayalite, for which the amorphous cross-section is not available \citep[see also][and \citetalias{Rogantini19}]{Zeegers17}.\\
However, if we are observing a real overabundance of crystalline grains, this apparent discrepancy with the infrared results might be attributed to the presence of poly-mineralic silicates which are expected to be agglomerated particles, possibly containing both glassy and crystalline constituents \citep{Marra11,Speck11}. In this case, because X-rays are sensitive to a short range order \citep{Mastelaro18}, XAFS would show crystalline features, whereas there might not be sharp crystalline features in the infrared spectrum \citep[see][ \citetalias{Rogantini19} and reference therein]{Zeegers19}. Infrared vibrational spectroscopy is indeed sensitive to a long-range order of the particles which is missing for poly-mineralic grains \citep{Oberti07}.\\
Although infrared and X-ray observations sample different lines of sight, they investigate the common dense medium towards the Galactic Centre, with comparable extinction values \cite[e.g][]{Kemper04,Li07,Min07}. Future dedicated studies will be necessary to reduce the uncertainties on the percentage of crystalline dust retrieved through infrared and X-ray spectroscopy.\\

\subsubsection{Olivine, pyroxene, quartz}

   \begin{figure*}
   \centering
     \includegraphics[width=0.95\hsize]{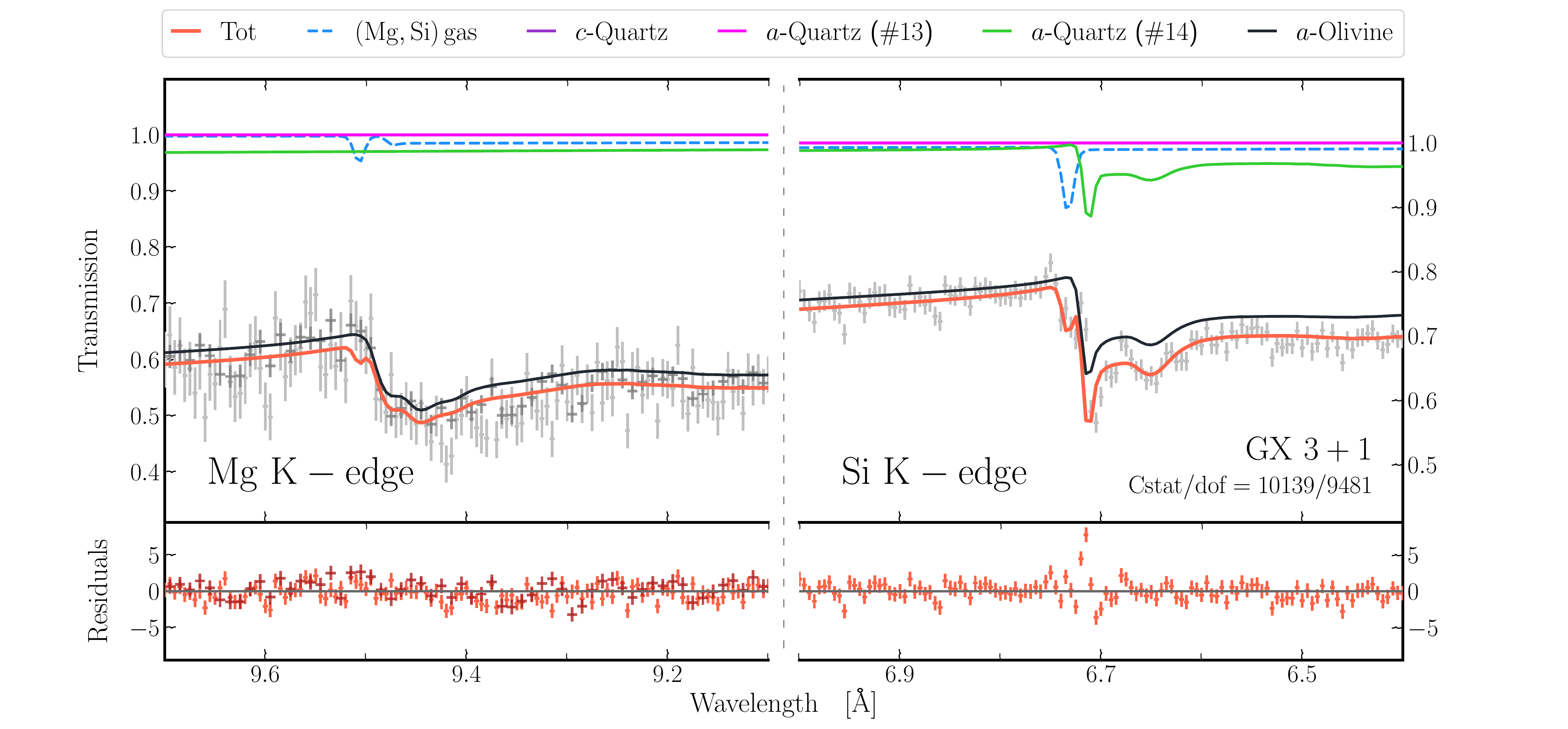}
      \caption{Zoom-in on the magnesium (on the left column) and silicon K-edge (on the right column) for GX 3+1. The HEG and MEG data are respectively shown in light and dark grey. We show the best fit obtained in the analysis together with contribution of each dust compound listed in the legend. The light-blue dashed line represents the absorption by magnesium and silicon in gas phase. The dust mixture is dominated by the amorphous olivine (see also Table \ref{tab:dust_properties}).}
         \label{fig:good_gx}
   \end{figure*}

   \begin{figure}
   \centering
     \includegraphics[width=0.97\hsize]{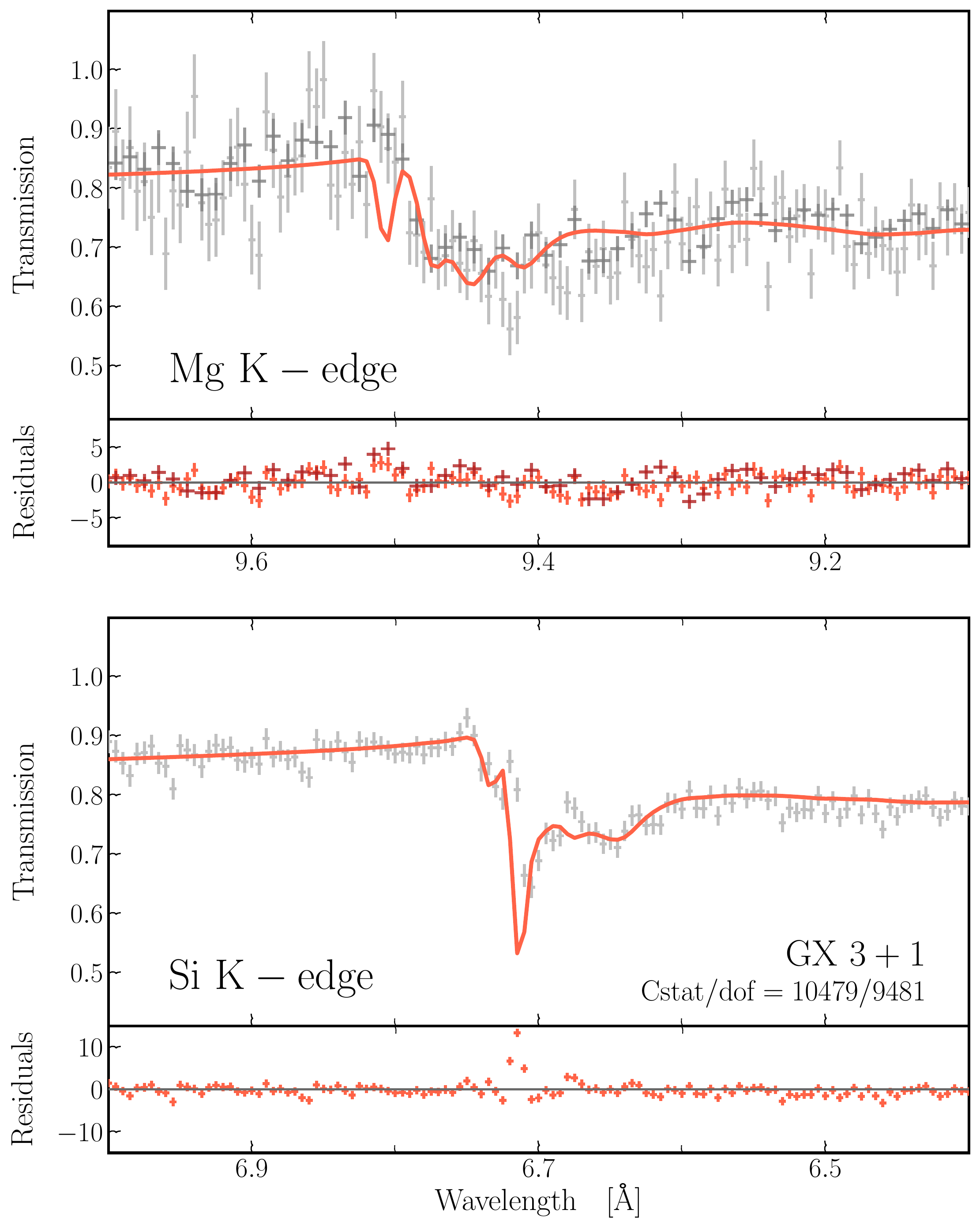}
      \caption{Zoom-in on the magnesium (on the left column) and silicon K-edge (on the right column) for GX 3+1. The HEG and MEG data are respectively shown in light and dark grey. We show the fit of a dust mixture without any olivinic compounds. In particular the dust counterpart consists of $c$-enstatite ($\sim 35\%$), $c$-En60Fs40 ($\sim 15\%$), $a$-quartz ($\sim 50\%$), and $c$-quartz ($<1\%$). Note the large residuals (up to $\sim 15\sigma$) in both edges.}
         \label{fig:bad_gx}
   \end{figure}

The dust observed along the line of sight of our sources shows overall an olivine structure characterised by the anion $[\rm SiO_4]^{-4}$. Indeed, we find values of \zii very close to unity (meaning that the dust is olivine-dominated) except for GX 9+1, for which we find a lower limit of 0.7 due to the large uncertainties on the column densities of the dust compounds. For all the considered fits pyroxene is not important. \\
To illustrate this, in Figure \ref{fig:good_gx} we display the importance of the olivine presence in the dust composition. The Mg and Si K-edges are shown in in transmission\footnote{Here, we consider the transmission for magnesium and silicon in both dust and gas. Thus, the observed counts are divided by the underlying continuum together with the cold absorption, except the two elements of interest.}, and together with the best fit (red solid line) we also plot the transmittance of each dust compounds (solid lines) and the transmittance of magnesium and silicon in gaseous form (dashed lines).\\
In Figure \ref{fig:bad_gx} we show an unacceptable fit obtained using a model which does not include any compound with an olivinic-configuration. Large residuals in both edges characterise the fit of dust mixtures without olivine.\\
Our results are consistent with \cite{Zeegers19}; they observe similarly a large fraction of olivine dust along the lines of sight of their sample of X-ray binaries near the Galactic centre. In the infrared, the broad silicate features are typically modelled with a mixture of olivines and pyroxenes \citep{Molster02,Chiar06,Henning10}. For example, \cite{Kemper04} found that cosmic silicates are composed of $\sim 85\%$ of olivine and $\sim15\%$ of pyroxene, values similar to our observations. \\ 
Differently, \cite{Fogerty16} studying the line of sight of heavily reddened stars observed with \spitzer found the presence of polivene, an amorphous silicate with intermediate stoichiometry \citep[$\rm Mg_{1.5}SiO_{3.5}$,][]{Jager03b}. This compound is not present in our sample set and it is not available in the literature. Ab initio methods \citep{Ramachandran08} do not provide enough accurate XAFS profile suitable for the purpose of our analysis. We plan to include the XAFS cross-section of compounds with a pyrosilicate anion ($\rm Si_2 O_7^{-6}$), in future X-ray studies. \\
Finally, the overall contribution of quartz in the fits of the sources is modest. Even in the fits with $\daic<10$, its fraction is always less than 10\%. This is consistent with the infrared spectroscopy results, where a strong absorption band by quartz particles in the interstellar medium has never been observed \citep{Sargent09}.

\subsection{Silicate cations}
Analysing the silicate features in the infrared spectrum of an evolved star, \cite{Kemper02} found that $\sim 80\%$ of the silicate is characterised by a comparable amount of magnesium and iron cations and less than $10\%$ is represented by magnesium-pure pyroxene ($\rm MgSiO_{3}$) and olivine ($\rm Mg_{2}SiO_{4}$). Similar results were found by \cite{Kemper04}, where they concluded that the cation composition of the dust in the interstellar medium is most likely $\ziii\approx 0.5-0.6$, where \ziii expresses the magnesium percentage overall cations (Equation \ref{eq:zetas}). This is in agreement with our results, where we find $\ziii \sim 0.5$ (Table \ref{tab:dust_properties}). Only two sources, H 1743-322 and GRS 1915+105, diverge from the average trend showing a preference for iron-rich olivine, namely fayalite. Whereas for the former source \ziii is affected by large uncertainties because of the quality of the data, for the latter it shows a significant deviation from the average value found in the other sources (Figure \ref{fig:aic_plot}).\\
The large amount of olivine observed along the line of sight of our X-ray binaries has important implications on the production and growth of interstellar dust in the Galaxy. Magnesium, iron and silicon are indeed present in olivine with the stoichiometric ratio 1:1:1. These elements share very similar abundance values and they are equally highly depleted in dust in the diffuse and cold interstellar medium \citep{Whittet02}. However, in the interstellar medium iron presents some important differences with respect to the other two elements. For example, it is heavily depleted also in the warm neutral medium following, thus, a distinct depletion behaviour \citep{Jenkins09}. Moreover, whereas magnesium and silicon condense onto silicate grains in the envelope of AGB stars and in the expanding core-collapse supernovae ejecta \citep{Hofner18,Dwek19} iron is mainly produced in Type Ia supernovae. In this environment, a dust envelope has proved to be challenging to observe \citep{Gomez12}. It is therefore believed that Fe is returned to the interstellar medium in gaseous form and the observed Fe depletion can be explained only by accretion of gas-phase Fe on grains \citep{Dwek16,Zhukovska18}. From the fit of the Mg and Si edges and the related \ziii values, we can extrapolate that a large fraction of iron is expected to be in interstellar silicates, at least for the considered lines of sight. The solid form of the interstellar iron is still highly debated and among the forms of iron considered in literature there are oxides \citep{Draine13}, metallic Fe \citep{Westphal19}, inclusions of Fe and FeS in silicate grains, free-flying iron particles \citep{Zhukovska18} and hydrogenated iron nanoparticles \citep{Bilalbegovic17}.\\
Future instruments, such as the coming X-ray Imaging and Spectroscopy Mission \citep[\textsc{XRISM,}][]{Tashiro18} and the future Athena X-ray mission \citep{Nandra13,Barret16}, will allow to simultaneously study the Mg and Si K-edges with the Fe K-edge, especially for high column-density lines of sight \citep{Rogantini18}. This will be crucial to disentangle the role of iron in the ISM.\\

\subsection{Abundances and depletions}
\label{sec:abund_depl}

\begin{table*}
\caption{Abundances and depletions. 
        }             
\label{tab:abu_dep}
\centering
\renewcommand{\arraystretch}{1.1}
\begin{tabular}{r @{\hspace{10mm}} c c c c c }       
\hline\hline
\noalign{\vskip 0.5mm}
  \multicolumn{1}{c}{\multirow{2}{*}{Source}} & $N^{\rm tot}_{\rm X}$ & $\delta_{\rm X}$ & $A_{\rm X}$ &  $A^{\rm dust}_{\rm X}$ & $A_{\rm X}/A_{\odot}$ \\
 & $\rm 10^{17}\,cm^{-2}$ & & $\rm 10^{-5}\,H^{-1}$ & $\rm 10^{-5}\,H^{-1}$ & \\
\noalign{\vskip 0.5mm}
\hline
\noalign{\vskip 0.5mm}
\rowcolor{Gray}              
 \multicolumn{6}{c}{Magnesium} \\
\hline
  GRS 1758-258    & $7.7\pm0.5$ & $>0.97$     & $3.6 \pm 0.3$  & $3.6 \pm 0.3$ & $0.9 \pm 0.1$        \\  
  GRS 1915+105    & $20\pm3   $ & $0.8\pm0.1$ & $4.0\pm0.5$    & $3.0\pm0.2$   & $1.0\pm0.1$          \\ 
  GX 17+2         & $9.4\pm2.2$ & $>0.97$     & $4.3\pm0.9$    & $4.3\pm0.9$   & $1.0\pm0.3$          \\
  GX 3+1          & $7.0\pm0.1$ & $>0.99$     & $3.7\pm0.1$    & $3.7\pm0.1$   & $0.9\pm0.1$          \\
  GX 9+1          & $8.5\pm3.5$ & $>0.90$     & $5\pm2$        & $5\pm2$       & $1.3\pm0.5$          \\
  H1743-322       & $9.6\pm2.8$ & $>0.74$     & $4\pm1$        & $3.0\pm0.7$   & $0.9\pm0.3$          \\
  IGR J17091-3624 & $4.4\pm0.2$ & $>0.96$     & $3.6\pm0.2$    & $3.6\pm0.2$   & $0.9\pm0.1$          \\
  LMC X-1         & $1.4\pm0.3$ & $>0.99$     & $0.8\pm0.2$    & $0.7\pm0.2$   & $0.20\pm0.05$        \\          
\noalign{\vskip 0.5mm}
  \hline               
  \rowcolor{Gray}      
 \multicolumn{6}{c}{Silicon} \\
\hline                      
\noalign{\vskip 0.5mm}
  GRS 1758-258    & $8.4\pm0.8$ & $0.98\pm0.01$   & $3.9\pm0.4$ & $3.8 \pm 0.4$ & $1.0 \pm 0.1$   \\
  GRS 1915+105    & $22\pm2$    & $0.98\pm0.01$   & $4.2\pm0.3$ & $4.2\pm0.3$   & $1.1 \pm0.1$    \\
  GX 17+2         & $9.3\pm2.0$ & $0.98\pm0.01$   & $4.3\pm0.9$ & $4.2\pm0.9$   & $1.1\pm0.2$     \\
  GX 3+1          & $8.7\pm0.2$ & $0.96\pm0.01$   & $4.5\pm0.1$ & $4.4\pm0.1$   & $1.2\pm0.1$     \\
  GX 9+1          & $7.7\pm2.3$ & $0.96 \pm 0.03$ & $5\pm1$     & $5\pm1$       & $1.2\pm0.4$     \\
  H1743-322       & $11\pm3$    & $0.97\pm0.02$   & $4\pm1$     & $4\pm1$       & $1.0\pm0.3$     \\
  IGR J17091-3624 & $4.8\pm0.3$ & $0.98\pm0.01$   & $4.0\pm0.3$ & $3.9\pm0.3$   & $1.0\pm0.1$     \\
  LMC X-1         & $0.9\pm0.3$ & $>0.98$         & $0.5\pm0.1$ & $0.5\pm0.1$   & $0.13\pm0.04$   \\       
\noalign{\vskip 0.5mm}
\hline                           
\end{tabular}
\tablefoot{$N_{X}^{\rm tot}$ represents the total column density of the X element, $\delta_{\rm X}$ its depletion from the gas phase, $A_{X}$ the total abundance, $A^{\rm dust}_{X}$ the abundance of the X element in dust. In Section \ref{sec:lmcx-1} we provide for LMC X-1 the abundance ratios obtained assuming the Large Magellanic Cloud abundances listed in Table \ref{tab:lmc_abundance}.}
\end{table*}

In Table \ref{tab:abu_dep}, we report both the depletion and abundance values of magnesium and silicon that we observed along the lines of sight of our X-ray binaries. We find that both elements are significantly depleted from the gas phase of the interstellar medium near the Galactic centre. Silicon, in particular, shows depletion values consistently higher than $0.95$. Since the signal-to-noise in the magnesium edge is usually lower than the silicon region, the uncertainties on the estimates are consequently larger, and we often find lower limits. These depletion values are consistent with \cite{Whittet02} and \cite{Jenkins09}.\\
Absolute abundances, calculated summing the gas and the dust contribution, show that in the environment near the Galactic centre, sampled by these sources, values are consistent with solar.\\
The abundances and depletion do not show any trend with the hydrogen column density. This is not surprising as our sources show a limited interval of line-of-sight column densities, ranging between $(1-5)\ \times 10^{22}\rm cm^{-2}$, which we can extend to $(1-10) \times 10^{22}\ \rm cm^{-2}$ for Si considering the sources studied by \cite{Zeegers19}. Moreover, our sources are all located within 5 kpc from the Galactic centre. In this view, our results on the abundances are in agreement with the results obtained by previous works \citep{Davies09,Genovali15,Martin15} which advance the idea that the abundance gradient of the magnesium and silicon flatten close to solar abundances in the inner part of the Galaxy.

\subsection{LMC X-1}
\label{sec:lmcx-1}

LMC X-1 is the only extragalactic source in our sample. It is the brightest X-ray source in the Large Magellanic Cloud, located at a distance of 48 kpc. Despite the long exposure time ($\sim160\ \rm ks$, see Table \ref{tab:sample}), the flux is not high enough to guarantee an optimal signal-to-noise ratio. Therefore, the best value of the fits are affected by larger uncertainties compared to the fits of the spectra of the Galactic sources.\\
Less than $12\%$ of the hydrogen column density towards the Large Magellanic Cloud, $4\times10^{21}\ \rm cm^{-2}$ \citep{Kalberla05}, is of Galactic origin \citep{Hanke10}. Thus, to model the absorption, we use the typical Large Magellanic Cloud abundances found in the literature and given in Table \ref{tab:lmc_abundance}. Similar to the Galactic lines of sight, magnesium and silicon are highly depleted from the gas phase. In particular, we observe depletion values for magnesium and silicon of $\delta_{\rm{Mg}} > 0.98 $ and $\delta_{\rm{Si}} > 0.99$, respectively. These results are consistent with the work by \cite{Tchernyshyov15}. \\
Comparing the abundances of the two elements with the abundances tabulated in Table \ref{tab:lmc_abundance} we find $A_{\rm Mg}/A_{\rm LMC} = 1.0\pm0.3 $ and $A_{\rm Si}/A_{\rm LMC} = 0.5\pm0.2$.  Moreover, the fit led by the Mg K-edge, tentatively show the presence of magnesium-rich silicate, in particular, the presence of forsterite which is the most representative compound in the dust mixture. We obtain a cation ratio \ziii with an upper limit of 0.76.  Since our set of compounds does not include the amorphous counterpart of forsterite, the crystallinity ratio, \zi, could be biased (see the discussion over GRS 1915+105 in Section \ref{sec:crystallinity} and in \citetalias{Rogantini19}). A possible presence of forsterite in its amorphous state would, therefore, decrease the crystalline ratio. \\
Our results hint at a possible under-abundance of silicon along the line of sight of LMC X-1. An abundance lower than the average is also found by \cite{Schenck16} in the two supernova remnants (0540-69.3 and DEM L316B) close to LMC X-1. We caution however that the quality of the data in the silicon region (see Figure \ref{fig:edges2}) affect our estimation. The values obtained studying the supernova remnants are also affected by large uncertainties \citep{Maggi16,Schenck16}. Moreover, the interstellar matter abundances may vary over the entire Magellanic Cloud.\\

\subsection{Two-edges fit}
\label{sec:two_edges}

In this work we made use of the simultaneous fit of two edges (magnesium and silicon) in order to derive the dust properties. This is important to remove some of the degeneracies in the model. For example, \cite{Zeegers19} only studied the Si K-edge of the GRS 1758-258 and GX 17+2 with the same technique and dust models used in the present work. Whereas for GX 17+2, which benefits from a high quality data, we find similar results when we include the magnesium K-edge, for GRS 1758-258 the joint fit of the two dust edges improves significantly the modelling of the absorption. Using a single edge, a global fit would not provide a definitive answer on the dust chemistry \citep{Zeegers19}.
However, thanks to the magnesium K-edge we can constrain the dust composition. The fit shows a striking preference for the amorphous olivine. Also the estimates of the abundances and depletion benefit from the addition of the Mg K-edge, decreasing the uncertainties on the relative best values by a factor of 4. The simultaneous fit of multiple edges helps to reduce the degeneracies of the fit and to better constrain the properties of the dust. This has been done, previously, by \mbox{\cite{Costantini12}} and \cite{Pinto13}, where they fit simultaneously the iron L-edges with the oxygen K-edge both located below 1 keV.

\section{Conclusions}
\label{sec:conclusion}
In this paper, we characterise the absorption by the material present in the dense environments towards and near the Galactic centre. We study the X-ray spectrum of seven bright X-ray binaries which lie in the Galactic plane. In particular, we inspect the magnesium and silicon K-edges at 1.3 and 1.84 keV, respectively. We evaluate the abundances of the two elements considering both dust and gas contributions. For every line of sight, we observe Mg and Si abundances consistent with the solar ones. Moreover, magnesium and silicon are highly depleted from the gas phase: we find that more than 90\% and 95\% of Mg and Si, respectively, are locked in dust grains.\\
Therefore, the interstellar dust largely contributes to shape the magnesium and silicon edges, making these two features a good probe to investigate the properties of the cosmic grains. We model them using accurate extinction models of silicates based on our laboratory measurements performed at the Soleil-LUCIA beamline. We conclude that:
\begin{itemize}
\item[-] \textit{dust composition}: for every line of sight, a high percentage of dust $(65-85)\%$ is represented by $\rm MgFeSiO_4$, an olivine with cation ratio $\ziii=0.5$. Magnesium-rich silicates are not preferred in the fits. Only the modelling of GRS 1915+105 possibly points to a prevalence of a more iron-rich olivine.
\item[-] \textit{dust crystallinity}: although the dust features are fitted mostly by dust with an amorphous structure, for most of the studied lines of sight the crystalline ratio (with an average of $\langle \zi \rangle = 11\%$) is larger than the ratio found in the infrared. For the outlier GRS 1915+105 the estimation of the percentage of crystalline dust is possibly affected by bias due to the incompleteness of our dust model set.
\item[-] \textit{dust structure}: in almost all the sources the olivine dust type $\rm [SiO_{4}]^{-2}$ is the dust mixture that best fit both Mg and Si K-edges. We do not find any significant presence of pyroxene along the studied lines of sight. Finally, the contribution of the quartz is always below $10\%$.
\end{itemize}
\noindent
In addition to the Galactic X-ray binaries, we explore the absorbed spectrum of the brightest source in the Large Magellanic Cloud, LMC X-1. Similar to the other source in the sample, Mg and Si are highly depleted. Si appears to be under-abundant along this line of sight, with respect to the average Large Magellanic Cloud abundances. The Si abundance is consistent with what found for neighbouring objects in the LMC. However, the results are affected by large uncertainties due to the poor photon statistics.

\begin{acknowledgements}
DR, EC, IP and MM are supported by the Netherlands Organisation for Scientific Research (NWO) through \emph{The Innovational Research Incentives Scheme Vidi} grant 639.042.525. The Space Research Organization of the Netherlands is financially supported by NWO. This research has made use of data obtained from the \emph{Chandra} Data Archive and the \emph{Chandra} Source Catalog, and software provided by the \emph{Chandra} X-ray Center (CXC) in the application package CIAO. This research made use of the Chandra Transmission Grating Catalog and archive (\url{http://tgcat.mit.edu}). We are grateful to H. Marshall for useful discussion on HETGS calibration. We thank M. Sasaki for discussion on the LMC X-1. We also thank A. Dekker and D. Lena for commenting on this manuscript. 
\end{acknowledgements}

\bibliographystyle{aa} 
\bibliography{../bibliography/biblio}

\begin{thebibliography}{118}
\expandafter\ifx\csname natexlab\endcsname\relax\def\natexlab#1{#1}\fi

\bibitem[{{Akaike}(1974)}]{Akaike74}
{Akaike}, H. 1974, IEEE Transactions on Automatic Control, 19, 716

\bibitem[{Akaike(1998)}]{Akaike98}
Akaike, H. 1998, Prediction and Entropy, ed. E.~Parzen, K.~Tanabe, \&
  G.~Kitagawa (New York, NY: Springer New York), 387--410

\bibitem[{{Arnett} \& {Thielemann}(1985)}]{Arnett85}
{Arnett}, W.~D. \& {Thielemann}, F.~K. 1985, \apj, 295, 589

\bibitem[{{Barret} {et~al.}(2016){Barret}, {Lam Trong}, {den Herder}, {Piro},
  {Barcons}, {Huovelin}, {Kelley}, {Mas-Hesse}, {Mitsuda}, {Paltani}, {Rauw},
  {Ro{\.Z}anska}, {Wilms}, {Barbera}, {Bozzo}, {Ceballos}, {Charles},
  {Decourchelle}, {den Hartog}, {Duval}, {Fiore}, {Gatti}, {Goldwurm},
  {Jackson}, {Jonker}, {Kilbourne}, {Macculi}, {Mendez}, {Molendi},
  {Orleanski}, {Pajot}, {Pointecouteau}, {Porter}, {Pratt}, {Pr{\^e}le},
  {Ravera}, {Renotte}, {Schaye}, {Shinozaki}, {Valenziano}, {Vink}, {Webb},
  {Yamasaki}, {Delcelier-Douchin}, {Le Du}, {Mesnager}, {Pradines},
  {Branduardi-Raymont}, {Dadina}, {Finoguenov}, {Fukazawa}, {Janiuk}, {Miller},
  {Naz{\'e}}, {Nicastro}, {Sciortino}, {Torrejon}, {Geoffray}, {Hernandez},
  {Luno}, {Peille}, {Andr{\'e}}, {Daniel}, {Etcheverry}, {Gloaguen}, {Hassin},
  {Hervet}, {Maussang}, {Moueza}, {Paillet}, {Vella}, {Campos Garrido},
  {Damery}, {Panem}, {Panh}, {Bandler}, {Biffi}, {Boyce}, {Cl{\'e}net},
  {DiPirro}, {Jamotton}, {Lotti}, {Schwander}, {Smith}, {van Leeuwen}, {van
  Weers}, {Brand}, {Cobo}, {Dauser}, {de Plaa}, \& {Cucchetti}}]{Barret16}
{Barret}, D., {Lam Trong}, T., {den Herder}, J.-W., {et~al.} 2016, in
  \procspie, Vol. 9905, Space Telescopes and Instrumentation 2016: Ultraviolet
  to Gamma Ray, 99052F

\bibitem[{{Berrington} {et~al.}(1995){Berrington}, {Eissner}, \&
  {Norrington}}]{Berrington95}
{Berrington}, K.~A., {Eissner}, W.~B., \& {Norrington}, P.~H. 1995, Computer
  Physics Communications, 92, 290

\bibitem[{{Bilalbegovi{\'c}} {et~al.}(2017){Bilalbegovi{\'c}},
  {Maksimovi{\'c}}, \& {Moha{\v{c}}ek-Gro{\v{s}}ev}}]{Bilalbegovic17}
{Bilalbegovi{\'c}}, G., {Maksimovi{\'c}}, A., \& {Moha{\v{c}}ek-Gro{\v{s}}ev},
  V. 2017, \mnras, 466, L14

\bibitem[{{Bilalbegovi{\'c}} {et~al.}(2018){Bilalbegovi{\'c}},
  {Maksimovi{\'c}}, \& {Valencic}}]{Bilalbegovic18}
{Bilalbegovi{\'c}}, G., {Maksimovi{\'c}}, A., \& {Valencic}, L.~A. 2018,
  \mnras, 476, 5358

\bibitem[{Burnham \& Anderson(2002)}]{Burnham02}
Burnham, K.~P. \& Anderson, D.~R. 2002, {Model selection and multimodel
  inference: a practical information-theoretic approach}, 2nd edn. (Springer),
  1--488

\bibitem[{{Cash}(1979)}]{Cash79}
{Cash}, W. 1979, \apj, 228, 939

\bibitem[{{Chiar} \& {Tielens}(2006)}]{Chiar06}
{Chiar}, J.~E. \& {Tielens}, A.~G.~G.~M. 2006, \apj, 637, 774

\bibitem[{{Corrales} {et~al.}(2016){Corrales}, {Garc{\'\i}a}, {Wilms}, \&
  {Baganoff}}]{Corrales16}
{Corrales}, L.~R., {Garc{\'\i}a}, J., {Wilms}, J., \& {Baganoff}, F. 2016,
  \mnras, 458, 1345

\bibitem[{{Costantini} {et~al.}(2005){Costantini}, {Freyberg}, \&
  {Predehl}}]{Costantini05}
{Costantini}, E., {Freyberg}, M.~J., \& {Predehl}, P. 2005, \aap, 444, 187

\bibitem[{{Costantini} {et~al.}(2012){Costantini}, {Pinto}, {Kaastra}, {in't
  Zand}, {Freyberg}, {Kuiper}, {M{\'e}ndez}, {de Vries}, \&
  {Waters}}]{Costantini12}
{Costantini}, E., {Pinto}, C., {Kaastra}, J.~S., {et~al.} 2012, \aap, 539, A32

\bibitem[{{Costantini} {et~al.}(2019){Costantini}, {Zeegers}, {Rogantini}, {de
  Vries}, {Tielens}, \& {Waters}}]{Costantini19}
{Costantini}, E., {Zeegers}, S.~T., {Rogantini}, D., {et~al.} 2019, \aap, 629,
  A78

\bibitem[{{Court} {et~al.}(2017){Court}, {Altamirano}, {Pereyra}, {Boon},
  {Yamaoka}, {Belloni}, {Wijnands}, \& {Pahari}}]{Court17}
{Court}, J.~M.~C., {Altamirano}, D., {Pereyra}, M., {et~al.} 2017, \mnras, 468,
  4748

\bibitem[{Cowan(1995)}]{Cowan95}
Cowan, J. 1995, The Biological Chemistry of Magnesium (Wiley)

\bibitem[{{Cowan}(1981)}]{Cowan81}
{Cowan}, R.~D. 1981, {The theory of atomic structure and spectra}

\bibitem[{{Davies} {et~al.}(2009){Davies}, {Origlia}, {Kudritzki}, {Figer},
  {Rich}, \& {Najarro}}]{Davies09}
{Davies}, B., {Origlia}, L., {Kudritzki}, R.-P., {et~al.} 2009, \apj, 694, 46

\bibitem[{{de Plaa} {et~al.}(2004){de Plaa}, {Kaastra}, {Tamura},
  {Pointecouteau}, {Mendez}, \& {Peterson}}]{dePlaa04}
{de Plaa}, J., {Kaastra}, J.~S., {Tamura}, T., {et~al.} 2004, \aap, 423, 49

\bibitem[{{den Hartog} {et~al.}(2003){den Hartog}, {in't Zand}, {Kuulkers},
  {Cornelisse}, {Heise}, {Bazzano}, {Cocchi}, {Natalucci}, \&
  {Ubertini}}]{denHartog03}
{den Hartog}, P.~R., {in't Zand}, J.~J.~M., {Kuulkers}, E., {et~al.} 2003,
  \aap, 400, 633

\bibitem[{{Draine}(2003)}]{Draine03}
{Draine}, B.~T. 2003, \apj, 598, 1026

\bibitem[{{Draine} \& {Hensley}(2013)}]{Draine13}
{Draine}, B.~T. \& {Hensley}, B. 2013, \apj, 765, 159

\bibitem[{{Dwek}(2016)}]{Dwek16}
{Dwek}, E. 2016, \apj, 825, 136

\bibitem[{Dwek {et~al.}(2019)Dwek, Sarangi, \& Arendt}]{Dwek19}
Dwek, E., Sarangi, A., \& Arendt, R.~G. 2019, The Astrophysical Journal, 871,
  L33

\bibitem[{{Egerton}(2009)}]{Egerton09}
{Egerton}, R.~F. 2009, Reports on Progress in Physics, 72, 016502

\bibitem[{{Fogerty} {et~al.}(2016){Fogerty}, {Forrest}, {Watson}, {Sargent}, \&
  {Koch}}]{Fogerty16}
{Fogerty}, S., {Forrest}, W., {Watson}, D.~M., {Sargent}, B.~A., \& {Koch}, I.
  2016, \apj, 830, 71

\bibitem[{{Fruscione} {et~al.}(2006){Fruscione}, {McDowell}, {Allen},
  {Brickhouse}, {Burke}, {Davis}, {Durham}, {Elvis}, {Galle}, {Harris},
  {Huenemoerder}, {Houck}, {Ishibashi}, {Karovska}, {Nicastro}, {Noble},
  {Nowak}, {Primini}, {Siemiginowska}, {Smith}, \& {Wise}}]{Fruscione06}
{Fruscione}, A., {McDowell}, J.~C., {Allen}, G.~E., {et~al.} 2006, in
  \procspie, Vol. 6270, Society of Photo-Optical Instrumentation Engineers
  (SPIE) Conference Series, 62701V

\bibitem[{{Galloway} {et~al.}(2008){Galloway}, {Muno}, {Hartman}, {Psaltis}, \&
  {Chakrabarty}}]{Galloway08}
{Galloway}, D.~K., {Muno}, M.~P., {Hartman}, J.~M., {Psaltis}, D., \&
  {Chakrabarty}, D. 2008, The Astrophysical Journal Supplement Series, 179, 360

\bibitem[{{Garnett}(1999)}]{Garnett99}
{Garnett}, D.~R. 1999, in IAU Symposium, Vol. 190, New Views of the Magellanic
  Clouds, ed. Y.-H. {Chu}, N.~{Suntzeff}, J.~{Hesser}, \& D.~{Bohlender}, 266

\bibitem[{{Genovali} {et~al.}(2014){Genovali}, {Lemasle}, {Bono}, {Romaniello},
  {Fabrizio}, {Ferraro}, {Iannicola}, {Laney}, {Nonino}, {Bergemann},
  {Buonanno}, {Fran{\c c}ois}, {Inno}, {Kudritzki}, {Matsunaga}, {Pedicelli},
  {Primas}, \& {Th{\'e}venin}}]{Genovali14}
{Genovali}, K., {Lemasle}, B., {Bono}, G., {et~al.} 2014, \aap, 566, A37

\bibitem[{{Genovali} {et~al.}(2015){Genovali}, {Lemasle}, {da Silva}, {Bono},
  {Fabrizio}, {Bergemann}, {Buonanno}, {Ferraro}, {Fran{\c{c}}ois},
  {Iannicola}, {Inno}, {Laney}, {Kudritzki}, {Matsunaga}, {Nonino}, {Primas},
  {Romaniello}, {Urbaneja}, \& {Th{\'e}venin}}]{Genovali15}
{Genovali}, K., {Lemasle}, B., {da Silva}, R., {et~al.} 2015, \aap, 580, A17

\bibitem[{{Gomez} {et~al.}(2012){Gomez}, {Clark}, {Nozawa}, {Krause}, {Gomez},
  {Matsuura}, {Barlow}, {Besel}, {Dunne}, {Gear}, {Hargrave}, {Henning},
  {Ivison}, {Sibthorpe}, {Swinyard}, \& {Wesson}}]{Gomez12}
{Gomez}, H.~L., {Clark}, C.~J.~R., {Nozawa}, T., {et~al.} 2012, \mnras, 420,
  3557

\bibitem[{Gu(2008)}]{Gu08}
Gu, M.~F. 2008, Canadian Journal of Physics, 86, 675

\bibitem[{{Hanke} {et~al.}(2010){Hanke}, {Wilms}, {Nowak}, {Barrag\'an}, \&
  {Schulz}}]{Hanke10}
{Hanke}, M., {Wilms}, J., {Nowak}, M.~A., {Barrag\'an}, L., \& {Schulz}, N.~S.
  2010, A\&A, 509, L8

\bibitem[{{Hasoglu} \& {Gorczyca}(2018)}]{Hasoglu18}
{Hasoglu}, M.~F. \& {Gorczyca}, T.~W. 2018, in Astronomical Society of the
  Pacific Conference Series, Vol. 515, Workshop on Astrophysical Opacities, 275

\bibitem[{{Henning}(2010)}]{Henning10}
{Henning}, T. 2010, \araa, 48, 21

\bibitem[{{Hoffman} \& {Draine}(2016)}]{Hoffman16}
{Hoffman}, J. \& {Draine}, B.~T. 2016, \apj, 817, 139

\bibitem[{H{\"o}fner \& Olofsson(2018)}]{Hofner18}
H{\"o}fner, S. \& Olofsson, H. 2018, The Astronomy and Astrophysics Review, 26,
  1

\bibitem[{{Honda} {et~al.}(2003){Honda}, {Kataza}, {Okamoto}, {Miyata},
  {Yamashita}, {Sako}, {Takubo}, \& {Onaka}}]{Honda03}
{Honda}, M., {Kataza}, H., {Okamoto}, Y.~K., {et~al.} 2003, \apjl, 585, L59

\bibitem[{{Huenemoerder} {et~al.}(2011){Huenemoerder}, {Mitschang}, {Dewey},
  {Nowak}, {Schulz}, {Nichols}, {Davis}, {Houck}, {Marshall}, {Noble},
  {Morgan}, \& {Canizares}}]{Huenemoerder11}
{Huenemoerder}, D.~P., {Mitschang}, A., {Dewey}, D., {et~al.} 2011, \aj, 141,
  129

\bibitem[{{Iaria} {et~al.}(2005){Iaria}, {di Salvo}, {Robba}, {Lavagetto},
  {Burderi}, {Stella}, \& {van der Klis}}]{Iaria05}
{Iaria}, R., {di Salvo}, T., {Robba}, N.~R., {et~al.} 2005, \aap, 439, 575

\bibitem[{{J{\"a}ger} {et~al.}(2003){J{\"a}ger}, {Dorschner}, {Mutschke},
  {Posch}, \& {Henning}}]{Jager03b}
{J{\"a}ger}, C., {Dorschner}, J., {Mutschke}, H., {Posch}, T., \& {Henning}, T.
  2003, \aap, 408, 193

\bibitem[{{Jenkins}(2009)}]{Jenkins09}
{Jenkins}, E.~B. 2009, \apj, 700, 1299

\bibitem[{{Jones}(2000)}]{Jones00}
{Jones}, A.~P. 2000, \jgr, 105, 10257

\bibitem[{Jones(2007)}]{Jones07}
Jones, A.~P. 2007, European Journal of Mineralogy, 19, 771

\bibitem[{{Jones} \& {Williams}(1987)}]{Jones87}
{Jones}, A.~P. \& {Williams}, D.~A. 1987, \mnras, 224, 473

\bibitem[{Jones(2015)}]{Jones15}
Jones, R.~O. 2015, Rev. Mod. Phys., 87, 897

\bibitem[{{Kaastra}(2017)}]{Kaastra17}
{Kaastra}, J.~S. 2017, \aap, 605, A51

\bibitem[{{Kaastra} \& {Barr}(1989)}]{Kaastra89}
{Kaastra}, J.~S. \& {Barr}, P. 1989, \aap, 226, 59

\bibitem[{{Kaastra} {et~al.}(1996){Kaastra}, {Mewe}, \&
  {Nieuwenhuijzen}}]{Kaastra96}
{Kaastra}, J.~S., {Mewe}, R., \& {Nieuwenhuijzen}, H. 1996, in UV and X-ray
  Spectroscopy of Astrophysical and Laboratory Plasmas, ed. K.~{Yamashita} \&
  T.~{Watanabe}, 411--414

\bibitem[{Kaastra {et~al.}(2018)Kaastra, Raassen, de~Plaa, \& Gu}]{Kaastra18}
Kaastra, J.~S., Raassen, A. J.~J., de~Plaa, J., \& Gu, L. 2018, SPEX X-ray
  spectral fitting package

\bibitem[{{Kalberla} {et~al.}(2005){Kalberla}, {Burton}, {Hartmann}, {Arnal},
  {Bajaja}, {Morras}, \& {P{\"o}ppel}}]{Kalberla05}
{Kalberla}, P.~M.~W., {Burton}, W.~B., {Hartmann}, D., {et~al.} 2005, \aap,
  440, 775

\bibitem[{{Kemper} {et~al.}(2002){Kemper}, {de Koter}, {Waters}, {Bouwman}, \&
  {Tielens}}]{Kemper02}
{Kemper}, F., {de Koter}, A., {Waters}, L.~B.~F.~M., {Bouwman}, J., \&
  {Tielens}, A.~G.~G.~M. 2002, \aap, 384, 585

\bibitem[{{Kemper} {et~al.}(2004){Kemper}, {Vriend}, \& {Tielens}}]{Kemper04}
{Kemper}, F., {Vriend}, W.~J., \& {Tielens}, A.~G.~G.~M. 2004, \apj, 609, 826

\bibitem[{{Kimura} {et~al.}(2003){Kimura}, {Mann}, \& {Jessberger}}]{Kimura03}
{Kimura}, H., {Mann}, I., \& {Jessberger}, E.~K. 2003, \apj, 583, 314

\bibitem[{Kirchhoff \& Bunsen(1860)}]{Kirchhoff60}
Kirchhoff, G. \& Bunsen, R. 1860, Annalen der Physik, 186, 161

\bibitem[{{Korn} {et~al.}(2000){Korn}, {Becker}, {Gummersbach}, \&
  {Wolf}}]{Korn00}
{Korn}, A.~J., {Becker}, S.~R., {Gummersbach}, C.~A., \& {Wolf}, B. 2000, \aap,
  353, 655

\bibitem[{{Lee} \& {Ravel}(2005)}]{Lee05}
{Lee}, J.~C. \& {Ravel}, B. 2005, \apj, 622, 970

\bibitem[{{Lee} {et~al.}(2002){Lee}, {Reynolds}, {Remillard}, {Schulz},
  {Blackman}, \& {Fabian}}]{Lee02}
{Lee}, J.~C., {Reynolds}, C.~S., {Remillard}, R., {et~al.} 2002, \apj, 567,
  1102

\bibitem[{{Lee} {et~al.}(2009){Lee}, {Xiang}, {Ravel}, {Kortright}, \&
  {Flanagan}}]{Lee09}
{Lee}, J.~C., {Xiang}, J., {Ravel}, B., {Kortright}, J., \& {Flanagan}, K.
  2009, \apj, 702, 970

\bibitem[{{Li} \& {Draine}(2001)}]{Li01a}
{Li}, A. \& {Draine}, B.~T. 2001, \apjl, 550, L213

\bibitem[{{Li} {et~al.}(2007){Li}, {Zhao}, \& {Li}}]{Li07}
{Li}, M.~P., {Zhao}, G., \& {Li}, A. 2007, \mnras, 382, L26

\bibitem[{{Lodders}(2010)}]{Lodders10}
{Lodders}, K. 2010, Astrophysics and Space Science Proceedings, 16, 379

\bibitem[{{Maggi} {et~al.}(2016){Maggi}, {Haberl}, {Kavanagh}, {Sasaki},
  {Bozzetto}, {Filipovi{\'c}}, {Vasilopoulos}, {Pietsch}, {Points}, {Chu},
  {Dickel}, {Ehle}, {Williams}, \& {Greiner}}]{Maggi16}
{Maggi}, P., {Haberl}, F., {Kavanagh}, P.~J., {et~al.} 2016, \aap, 585, A162

\bibitem[{{Marra} {et~al.}(2011){Marra}, {Lane}, {Orofino}, {Blanco}, \&
  {Fonti}}]{Marra11}
{Marra}, A.~C., {Lane}, M.~D., {Orofino}, V., {Blanco}, A., \& {Fonti}, S.
  2011, \icarus, 211, 839

\bibitem[{{Martin} {et~al.}(2015){Martin}, {Andrievsky}, {Kovtyukh}, {Korotin},
  {Yegorova}, \& {Saviane}}]{Martin15}
{Martin}, R.~P., {Andrievsky}, S.~M., {Kovtyukh}, V.~V., {et~al.} 2015, \mnras,
  449, 4071

\bibitem[{Mastelaro \& Zanotto(2018)}]{Mastelaro18}
Mastelaro, V.~R. \& Zanotto, E.~D. 2018, Materials, 11

\bibitem[{{Mathis} {et~al.}(1977){Mathis}, {Rumpl}, \& {Nordsieck}}]{Mathis77}
{Mathis}, J.~S., {Rumpl}, W., \& {Nordsieck}, K.~H. 1977, \apj, 217, 425

\bibitem[{{Mattsson} {et~al.}(2019){Mattsson}, {De Cia}, {Andersen}, \&
  {Petitjean}}]{Mattsson19}
{Mattsson}, L., {De Cia}, A., {Andersen}, A.~C., \& {Petitjean}, P. 2019, \aap,
  624, A103

\bibitem[{{Mehdipour} {et~al.}(2016){Mehdipour}, {Kaastra}, \&
  {Kallman}}]{Mehdipour16}
{Mehdipour}, M., {Kaastra}, J.~S., \& {Kallman}, T. 2016, \aap, 596, A65

\bibitem[{{Miller} {et~al.}(2002){Miller}, {Fabian}, {Wijnands}, {Remillard},
  {Wojdowski}, {Schulz}, {Di Matteo}, {Marshall}, {Canizares}, {Pooley}, \&
  {Lewin}}]{Miller02}
{Miller}, J.~M., {Fabian}, A.~C., {Wijnands}, R., {et~al.} 2002, \apj, 578, 348

\bibitem[{{Miller} {et~al.}(2005){Miller}, {Wojdowski}, {Schulz}, {Marshall},
  {Fabian}, {Remillard}, {Wijnand s}, \& {Lewin}}]{Miller05}
{Miller}, J.~M., {Wojdowski}, P., {Schulz}, N.~S., {et~al.} 2005, \apj, 620,
  398

\bibitem[{{Min} {et~al.}(2007){Min}, {Waters}, {de Koter}, {Hovenier},
  {Keller}, \& {Markwick-Kemper}}]{Min07}
{Min}, M., {Waters}, L.~B.~F.~M., {de Koter}, A., {et~al.} 2007, \aap, 462, 667

\bibitem[{{Mitsuda} {et~al.}(1984){Mitsuda}, {Inoue}, {Koyama}, {Makishima},
  {Matsuoka}, {Ogawara}, {Shibazaki}, {Suzuki}, {Tanaka}, \&
  {Hirano}}]{Mitsuda84}
{Mitsuda}, K., {Inoue}, H., {Koyama}, K., {et~al.} 1984, \pasj, 36, 741

\bibitem[{{Molster} {et~al.}(2010){Molster}, {Waters}, \& {Kemper}}]{Molster10}
{Molster}, F.~J., {Waters}, L.~B.~F.~M., \& {Kemper}, F. 2010, {The Mineralogy
  of Interstellar and Circumstellar Dust in Galaxies}, ed. T.~{Henning}, Vol.
  815, 143--201

\bibitem[{{Molster} {et~al.}(2002){Molster}, {Waters}, \&
  {Tielens}}]{Molster02}
{Molster}, F.~J., {Waters}, L.~B.~F.~M., \& {Tielens}, A.~G.~G.~M. 2002, \aap,
  382, 222

\bibitem[{{Nandra} {et~al.}(2013){Nandra}, {Barret}, {Barcons}, {Fabian}, {den
  Herder}, {Piro}, {Watson}, {Adami}, {Aird}, {Afonso}, \& et~al.}]{Nandra13}
{Nandra}, K., {Barret}, D., {Barcons}, X., {et~al.} 2013, ArXiv e-prints
  [\eprint[arXiv]{1306.2307}]

\bibitem[{{Natta} {et~al.}(2007){Natta}, {Testi}, {Calvet}, {Henning},
  {Waters}, \& {Wilner}}]{Natta07}
{Natta}, A., {Testi}, L., {Calvet}, N., {et~al.} 2007, in Protostars and
  Planets V, ed. B.~{Reipurth}, D.~{Jewitt}, \& K.~{Keil}, 767

\bibitem[{Newville(2004)}]{Newville04}
Newville, M. 2004, Consortium for Advanced Radiation Sources, University of
  Chicago (USA)[http://xafs. org], 78

\bibitem[{Oberti {et~al.}(2007)Oberti, Hawthorne, Cannillo, \&
  C{\'a}mara}]{Oberti07}
Oberti, R., Hawthorne, F., Cannillo, E., \& C{\'a}mara, F. 2007, Reviews in
  Mineralogy and Geochemistry, 67, 125

\bibitem[{{Orosz} {et~al.}(2009){Orosz}, {Steeghs}, {McClintock}, {Torres},
  {Bochkov}, {Gou}, {Narayan}, {Blaschak}, {Levine}, {Remillard}, {Bailyn},
  {Dwyer}, \& {Buxton}}]{Orosz09}
{Orosz}, J.~A., {Steeghs}, D., {McClintock}, J.~E., {et~al.} 2009, \apj, 697,
  573

\bibitem[{{Palme} {et~al.}(2014){Palme}, {Lodders}, \& {Jones}}]{Palme14}
{Palme}, H., {Lodders}, K., \& {Jones}, A. 2014, {Solar System Abundances of
  the Elements}, ed. A.~M. {Davis}, Vol.~2, 15--36

\bibitem[{{Pedicelli} {et~al.}(2009){Pedicelli}, {Bono}, {Lemasle}, {Fran{\c
  c}ois}, {Groenewegen}, {Lub}, {Pel}, {Laney}, {Piersimoni}, {Romaniello},
  {Buonanno}, {Caputo}, {Cassisi}, {Castelli}, {Leurini}, {Pietrinferni},
  {Primas}, \& {Pritchard}}]{Pedicelli09}
{Pedicelli}, S., {Bono}, G., {Lemasle}, B., {et~al.} 2009, \aap, 504, 81

\bibitem[{{Pinto} {et~al.}(2013){Pinto}, {Kaastra}, {Costantini}, \& {de
  Vries}}]{Pinto13}
{Pinto}, C., {Kaastra}, J.~S., {Costantini}, E., \& {de Vries}, C. 2013, \aap,
  551, A25

\bibitem[{{Pinto} {et~al.}(2010){Pinto}, {Kaastra}, {Costantini}, \&
  {Verbunt}}]{Pinto10}
{Pinto}, C., {Kaastra}, J.~S., {Costantini}, E., \& {Verbunt}, F. 2010, \aap,
  521, A79

\bibitem[{{Ramachandran} {et~al.}(2008){Ramachandran}, {Gopakumar}, \&
  {Namboori}}]{Ramachandran08}
{Ramachandran}, K.~I., {Gopakumar}, D., \& {Namboori}, K. 2008, The Ab Initio
  Method (Berlin, Heidelberg: Springer Berlin Heidelberg), 155--170

\bibitem[{{Reid} {et~al.}(2014){Reid}, {McClintock}, {Steiner}, {Steeghs},
  {Remillard}, {Dhawan}, \& {Narayan}}]{Reid14}
{Reid}, M.~J., {McClintock}, J.~E., {Steiner}, J.~F., {et~al.} 2014, \apj, 796,
  2

\bibitem[{{Rich} {et~al.}(2017){Rich}, {Ryde}, {Thorsbro}, {Fritz},
  {Schultheis}, {Origlia}, \& {J{\"o}nsson}}]{Rich17}
{Rich}, R.~M., {Ryde}, N., {Thorsbro}, B., {et~al.} 2017, \aj, 154, 239

\bibitem[{{Rogantini} {et~al.}(2018){Rogantini}, {Costantini}, {Zeegers}, {de
  Vries}, {Bras}, {de Groot}, {Mutschke}, \& {Waters}}]{Rogantini18}
{Rogantini}, D., {Costantini}, E., {Zeegers}, S.~T., {et~al.} 2018, \aap, 609,
  A22

\bibitem[{{Rogantini} {et~al.}(2019){Rogantini}, {Costantini}, {Zeegers}, {de
  Vries}, {Mehdipour}, {de Groot}, {Mutschke}, {Psaradaki}, \&
  {Waters}}]{Rogantini19}
{Rogantini}, D., {Costantini}, E., {Zeegers}, S.~T., {et~al.} 2019, \aap, 630,
  A143

\bibitem[{{Rolleston} {et~al.}(2000){Rolleston}, {Smartt}, {Dufton}, \&
  {Ryans}}]{Rolleston00}
{Rolleston}, W.~R.~J., {Smartt}, S.~J., {Dufton}, P.~L., \& {Ryans}, R.~S.~I.
  2000, \aap, 363, 537

\bibitem[{{Russell} \& {Dopita}(1992)}]{Russell92}
{Russell}, S.~C. \& {Dopita}, M.~A. 1992, \apj, 384, 508

\bibitem[{{Rybicki} {et~al.}(1986){Rybicki}, {Lightman}, \& {Paul}}]{Rybicki86}
{Rybicki}, G.~B., {Lightman}, A.~P., \& {Paul}, H.~G. 1986, Astronomische
  Nachrichten, 307, 170

\bibitem[{{Sargent} {et~al.}(2009){Sargent}, {Forrest}, {Tayrien}, {McClure},
  {Li}, {Basu}, {Manoj}, {Watson}, {Bohac}, {Furlan}, {Kim}, {Green}, \&
  {Sloan}}]{Sargent09}
{Sargent}, B.~A., {Forrest}, W.~J., {Tayrien}, C., {et~al.} 2009, \apj, 690,
  1193

\bibitem[{Schenck {et~al.}(2016)Schenck, Park, \& Post}]{Schenck16}
Schenck, A., Park, S., \& Post, S. 2016, The Astronomical Journal, 151, 161

\bibitem[{{Schultheis} {et~al.}(2019){Schultheis}, {Rich, R. M.}, {Origlia,
  L.}, {Ryde, N.}, {Nandakumar, G.}, {Thorsbro, B.}, \& {Neumayer,
  N.}}]{Schultheis19}
{Schultheis}, M., {Rich, R. M.}, {Origlia, L.}, {et~al.} 2019, A\&A, 627, A152

\bibitem[{{Schulz} {et~al.}(2016){Schulz}, {Corrales}, \&
  {Canizares}}]{Schulz16}
{Schulz}, N.~S., {Corrales}, L., \& {Canizares}, C.~R. 2016, in AAS/High Energy
  Astrophysics Division, Vol.~15, AAS/High Energy Astrophysics Division, 402.05

\bibitem[{{Shakura} \& {Sunyaev}(1973)}]{Shakura-Sunyaev73}
{Shakura}, N.~I. \& {Sunyaev}, R.~A. 1973, in IAU Symposium, Vol.~55, X- and
  Gamma-Ray Astronomy, ed. H.~{Bradt} \& R.~{Giacconi}, 155

\bibitem[{{Soria} {et~al.}(2011){Soria}, {Broderick}, {Hao}, {Hannikainen},
  {Mehdipour}, {Pottschmidt}, \& {Zhang}}]{Soria11}
{Soria}, R., {Broderick}, J.~W., {Hao}, J., {et~al.} 2011, \mnras, 415, 410

\bibitem[{{Speck} {et~al.}(2011){Speck}, {Whittington}, \&
  {Hofmeister}}]{Speck11}
{Speck}, A.~K., {Whittington}, A.~G., \& {Hofmeister}, A.~M. 2011, \apj, 740,
  93

\bibitem[{{Steenbrugge} {et~al.}(2005){Steenbrugge}, {Kaastra}, {Crenshaw},
  {Kraemer}, {Arav}, {George}, {Liedahl}, {van der Meer}, {Paerels}, {Turner},
  \& {Yaqoob}}]{Steenbrugge05}
{Steenbrugge}, K.~C., {Kaastra}, J.~S., {Crenshaw}, D.~M., {et~al.} 2005, \aap,
  434, 569

\bibitem[{{Steenbrugge} {et~al.}(2003){Steenbrugge}, {Kaastra}, {de Vries}, \&
  {Edelson}}]{Steenbrugge03}
{Steenbrugge}, K.~C., {Kaastra}, J.~S., {de Vries}, C.~P., \& {Edelson}, R.
  2003, \aap, 402, 477

\bibitem[{{Steiner} {et~al.}(2012){Steiner}, {McClintock}, \&
  {Reid}}]{Steiner12}
{Steiner}, J.~F., {McClintock}, J.~E., \& {Reid}, M.~J. 2012, \apjl, 745, L7

\bibitem[{{Takei} {et~al.}(2003){Takei}, {Fujimoto}, {Mitsuda}, {Futamoto}, \&
  {Onaka}}]{Takei03}
{Takei}, Y., {Fujimoto}, R., {Mitsuda}, K., {Futamoto}, K., \& {Onaka}, T.
  2003, in Astrophysics of Dust, 45

\bibitem[{{Tashiro} {et~al.}(2018){Tashiro}, {Maejima}, {Toda}, {Kelley},
  {Reichenthal}, {Lobell}, {Petre}, {Guainazzi}, {Costantini}, \&
  {Edison}}]{Tashiro18}
{Tashiro}, M., {Maejima}, H., {Toda}, K., {et~al.} 2018, in Society of
  Photo-Optical Instrumentation Engineers (SPIE) Conference Series, Vol. 10699,
  \procspie, 1069922

\bibitem[{{Tchernyshyov} {et~al.}(2015){Tchernyshyov}, {Meixner}, {Seale},
  {Fox}, {Friedman}, {Dwek}, \& {Galliano}}]{Tchernyshyov15}
{Tchernyshyov}, K., {Meixner}, M., {Seale}, J., {et~al.} 2015, \apj, 811, 78

\bibitem[{{Thielemann} \& {Arnett}(1985)}]{Thielemann85}
{Thielemann}, F.~K. \& {Arnett}, W.~D. 1985, \apj, 295, 604

\bibitem[{{Titarchuk}(1994)}]{Titarchuk94}
{Titarchuk}, L. 1994, \apj, 434, 570

\bibitem[{{Ueda} {et~al.}(2005){Ueda}, {Mitsuda}, {Murakami}, \&
  {Matsushita}}]{Ueda05}
{Ueda}, Y., {Mitsuda}, K., {Murakami}, H., \& {Matsushita}, K. 2005, \apj, 620,
  274

\bibitem[{{Ueda} {et~al.}(2009){Ueda}, {Yamaoka}, \& {Remillard}}]{Ueda09}
{Ueda}, Y., {Yamaoka}, K., \& {Remillard}, R. 2009, \apj, 695, 888

\bibitem[{{Valencic} {et~al.}(2009){Valencic}, {Smith}, {Dwek}, {Graessle}, \&
  {Dame}}]{Valencic09}
{Valencic}, L.~A., {Smith}, R.~K., {Dwek}, E., {Graessle}, D., \& {Dame}, T.~M.
  2009, \apj, 692, 502

\bibitem[{{Verner} {et~al.}(1993){Verner}, {Yakovlev}, {Band}, \&
  {Trzhaskovskaya}}]{Verner93}
{Verner}, D.~A., {Yakovlev}, D.~G., {Band}, I.~M., \& {Trzhaskovskaya}, M.~B.
  1993, Atomic Data and Nuclear Data Tables, 55, 233

\bibitem[{{Westphal} {et~al.}(2019){Westphal}, {Butterworth}, {Tomsick}, \&
  {Gainsforth}}]{Westphal19}
{Westphal}, A.~J., {Butterworth}, A.~L., {Tomsick}, J.~A., \& {Gainsforth}, Z.
  2019, \apj, 872, 66

\bibitem[{{Westphal} {et~al.}(2014){Westphal}, {Stroud}, {Bechtel}, {Brenker},
  {Butterworth}, {Flynn}, {Frank}, {Gainsforth}, {Hillier}, \&
  {Postberg}}]{Westphal14}
{Westphal}, A.~J., {Stroud}, R.~M., {Bechtel}, H.~A., {et~al.} 2014, Science,
  345, 786

\bibitem[{Whittet(2002)}]{Whittet02}
Whittet, D. 2002, Dust in the Galactic Environment, 2nd Edition, Series in
  Astronomy and Astrophysics (Taylor \& Francis)

\bibitem[{{Zeegers} {et~al.}(2017){Zeegers}, {Costantini}, {de Vries},
  {Tielens}, {Chihara}, {de Groot}, {Mutschke}, {Waters}, \&
  {Zeidler}}]{Zeegers17}
{Zeegers}, S.~T., {Costantini}, E., {de Vries}, C.~P., {et~al.} 2017, \aap,
  599, A117

\bibitem[{{Zeegers} {et~al.}(2019){Zeegers}, {Costantini}, {Rogantini}, {de
  Vries}, {Mutschke}, {Mohr}, {de Groot}, \& {Tielens}}]{Zeegers19}
{Zeegers}, S.~T., {Costantini}, E., {Rogantini}, D., {et~al.} 2019, \aap, 627,
  A16

\bibitem[{{Zhukovska} {et~al.}(2018){Zhukovska}, {Henning}, \&
  {Dobbs}}]{Zhukovska18}
{Zhukovska}, S., {Henning}, T., \& {Dobbs}, C. 2018, \apj, 857, 94

\end{thebibliography}

\begin{appendix}

\section{Broadband spectra}
\label{app:broadband}
We give an overview of the parameters obtained from the best fit modelling of the X-ray binaries studied this work. As mentioned in Section \ref{sec:source_selection}, the sources were selected by their value of \NH and flux.\\

\noindent
 The effect of pileup on these bright sources can be high and distort the signal from the source. Therefore, we ignore the regions of the spectra where this effect is present. In particular, MEG, for the shape of its effective area, is affected by pile-up around 2 keV. Thus, in many observations, we exclude MEG data around the region of the silicon K-edge. GRS 1915+105 (obsid 7458) presents an extreme case where the effect of pile-up is higher than 40\% around 2-3 keV because of the high 2-10 keV flux. For this specific source we only select the energy band between 1.2 and 2.4 keV of the HEG spectrum, whereas for MEG we ignore the region above 1.7 keV. In this way, we limited the spectrum into a narrow band which we fit adopting the pile-up corrected model used by \cite{Lee02}.\\

 \noindent
In general, we fit the continuum of the sources using both thermal and non-thermal components (see Section \ref{sec:continuum}) absorbed by interstellar matter. In Table \ref{tab:result_fit} we list all the models and the best values of their relative parameters. For each source, the hydrogen column density does not change significantly among the different observations and we assume it constant for all the observations. The Large Magellanic Cloud (LMC) has a much lower metallicity than our Galaxy \citep{Russell92}. We compile both abundances sets in Table \ref{tab:lmc_abundance}.\\

\noindent
Finally, we inspect the spectra for the presence of out-flowing ionised gas and hot gas present along the line of sight. Indeed, if absorption lines of ionised gas appear close to the edge, it is necessary to take them into account for precise modelling. We test whether if adding a \hot model (to model collisionally ionised gas), \xabs (to model photoionised gas), or a combination of the two improves the statistics of the broadband fit. As specified in Section \ref{sec:continuum} we find evidence of gas intrinsic to the source only in the microquasar GRS 1915+105.

%
\begin{sidewaystable*}
\caption{Best fit parameters for all the sources.}
\footnotesize
\label{tab:result_fit}
\centering
\begin{tabular}{c c c c c c c c c c c c c } 
\hline\hline
        
 & {\tt hot} & {\tt pow} & \multicolumn{2}{c}{\tt comt} & {\tt bb} & {\tt  dbb} & \multicolumn{3}{c}{\tt xabs} & &  & \\\cline{4-5}\cline{8-10}

\multirow{-2}{*}{obsID}       & $N_{\rm H}$ & $\Gamma $ & $k_{\rm B}T$ & $\tau$ & $k_{\rm B}T$ & $k_{\rm B}T$ & $N_{\rm H}$ & $\log \xi$ & v & \multirow{-2}{*}{$F_{0.5-2\rm\,keV}$} & \multirow{-2}{*}{$F_{2-10\rm\,keV}$} & \multirow{-2}{*}{Cstat/dof} \\

& $\rm 10^{22}\, cm^{-2}$ & & kev & & kev & kev &  $\rm 10^{20}\, cm^{-2}$ &  & $\rm 10^2\,km/s$ &  $\rm  10^{-10}\ erg\,cm^{-2}\,s^{-1}$ & $\rm  10^{-9}\ erg\,cm^{-2}\,s^{-1}$ & \\
\hline
\rowcolor{Gray}
\multicolumn{13}{c}{GRS 1758-258}\\
\hline
2429  & \multirow{2}{*}{$2.14 \pm 0.05$} & $3.9\pm0.2$ & -- & -- & $0.90\pm0.06$ & -- & -- & -- & -- & $0.79\pm0.08$ & $0.12\pm0.01$ & \multirow{2}{*}{$12621/11561$}\\
2750  & & $2.8\pm0.1$ & -- & -- & $1.00\pm0.01$ &  -- & -- & -- & -- & $1.2\pm0.2$ & $0.42\pm0.06$ & \\
\hline
\rowcolor{Gray}
\multicolumn{13}{c}{GRS 1915+105} \\
\hline
660   & \multirow{2}{*}{$5.06 \pm 0.02$} & -- & -- & -- & -- & $3.0\pm0.1$ & $3\pm1$ & $2.7\pm0.2$ & $-7\pm2$ & $1.5\pm0.1$ & $7.8\pm0.8$ & \multirow{2}{*}{$6056/5241$} \\
7485  & & -- & -- & -- & -- & $12\pm3$ & $5.9\pm0.7$ & $3.72\pm0.02$ & $-1.45\pm0.09$ & $1.3\pm0.4$ & $23\pm7$ & \\
\hline
\rowcolor{Gray}
\multicolumn{13}{c}{GX 3+1} \\
\hline
16492 & \multirow{7}{*}{$1.91 \pm 0.05$} & $1.12\pm0.04$ & -- & -- & $0.8\pm0.1$ & -- & -- & -- & -- & $3.7\pm0.2$ & $7.5\pm0.5$ & \multirow{7}{*}{28199/27207}\\
16307 & & $1.09\pm0.04$ & -- & -- & $0.8\pm0.1$ & -- & -- & -- & -- & $3.6\pm0.2$  & $7.4\pm0.5$ & \\
18615 & & $1.25\pm0.04$ & -- & -- & $0.7\pm0.2$ & -- & -- & -- & -- & $2.6\pm0.1$  & $4.8\pm0.3$ & \\
19890 & & $1.23\pm0.03$ & -- & -- & $0.8\pm0.2$ & -- & -- & -- & -- & $3.3\pm0.2$  & $6.1\pm0.4$ & \\
19907 & & $1.21\pm0.04$ & -- & -- & $0.7\pm0.1$ & -- & -- & -- & -- & $2.7\pm0.1$  & $4.8\pm0.3$ & \\
19957 & & $1.20\pm0.03$ & -- & -- & $0.8\pm0.2$ & -- & -- & -- & -- & $3.4\pm0.2$  & $6.9\pm0.4$ & \\
19958 & & $1.22\pm0.03$ & -- & -- & $0.8\pm0.1$ & -- & -- & -- & -- & $3.3\pm0.2$  & $6.2\pm0.4$ & \\
\hline
\rowcolor{Gray}
\multicolumn{13}{c}{GX 9+1} \\
\hline
11088 & $ 1.65 \pm 0.02  $ & -- & $2.4\pm0.4$ & $22\pm9$ & $0.76\pm0.04$ & -- & - & -- & -- & $0.53\pm0.04$ & $10\pm1$ & $4815/4293$ \\
\hline
\rowcolor{Gray}
\multicolumn{13}{c}{GX 17+2} \\
\hline
717   & $ 2.19 \pm 0.01  $ & -- & $2.5\pm0.1$ & $16\pm4$ & $0.79\pm0.02$ & -- & -- & -- & -- &$4.3\pm0.2$ & $13\pm1$ & 4946/4244 \\
\hline
\rowcolor{Gray}
\multicolumn{13}{c}{H 17433-322} \\
\hline
16738 & \multirow{4}{*}{$2.67 \pm 0.09$} & $1.47\pm0.04$ & -- & -- & -- & $0.38\pm0.09$ & -- & -- & -- & $0.35\pm0.02$ & $0.88\pm0.05$ & \multirow{4}{*}{$19169/18365$} \\
17679 & & $1.47\pm0.03$ & -- & -- & -- & $0.45\pm0.05$ & -- & -- & -- & $0.21\pm0.01$  & $0.54\pm0.03$ & \\
17680 & & $1.48\pm0.04$ & -- & -- & -- & $0.42\pm0.06$ & -- & -- & -- & $0.35\pm0.02$  & $0.91\pm0.05$ & \\
16739 & & $1.49\pm0.04$ & -- & -- & -- & $0.44\pm0.08$ & -- & -- & -- & $0.39\pm0.02$  & $0.96\pm0.05$ & \\
\hline
\rowcolor{Gray}
\multicolumn{13}{c}{IGR J17091-3624} \\
\hline
12406 & \multirow{3}{*}{$1.20 \pm 0.01$} & $1.78\pm0.03$ & -- & -- & $0.86\pm0.01$ & -- & -- & -- & -- & $2.5\pm0.1$ & $1.9\pm0.1$ & \multirow{3}{*}{$17403/16586$} \\
17787 & & $2.04\pm0.02$ & -- & -- & $0.55\pm0.01 $ & -- & -- & -- & -- & $2.3\pm0.2$ & $1.1\pm0.1$ & \\
17788 & & $1.99\pm0.03$ & -- & -- & $0.69\pm0.01 $ & -- & -- & -- & -- & $1.9\pm0.1$ & $1.0\pm0.1$ & \\
\hline
\rowcolor{Gray}
\multicolumn{13}{c}{LMC X-1} \\
\hline
93    &  \multirow{11}{*}{$0.79 \pm 0.01$} & $2.46\pm0.03$ & -- & -- & $0.61\pm0.01$ & -- & -- & -- & -- & $1.6\pm0.1$ & $0.38\pm0.03$ & \multirow{11}{*}{$22659/21818$}\\
11074 & & $2.45\pm0.04$ & -- & -- & $0.63\pm0.01$ & -- & -- & -- & -- & $1.7\pm0.1$ & $0.42\pm0.04$ & \\
11986 & & $2.65\pm0.05$ & -- & -- & $0.68\pm0.01$ & -- & -- & -- & -- & $1.4\pm0.1$ & $0.32\pm0.03$ & \\
11987 & & $2.44\pm0.04$ & -- & -- & $0.62\pm0.01$ & -- & -- & -- & -- & $1.6\pm0.1$ & $0.37\pm0.03$ & \\
12068 & & $2.40\pm0.03$ & -- & -- & $0.70\pm0.01$ & -- & -- & -- & -- & $1.7\pm0.1$ & $0.47\pm0.04$ & \\
12071 & & $2.59\pm0.05$ & -- & -- & $0.61\pm0.01$ & -- & -- & -- & -- & $1.8\pm0.1$ & $0.39\pm0.04$ & \\
12069 & & $2.51\pm0.07$ & -- & -- & $0.64\pm0.01$ & -- & -- & -- & -- & $1.7\pm0.1$ & $0.41\pm0.03$ & \\
12070 & & $2.62\pm0.06$ & -- & -- & $0.60\pm0.01$ & -- & -- & -- & -- & $1.6\pm0.1$ & $0.34\pm0.03$ & \\
12089 & & $2.58\pm0.05$ & -- & -- & $0.64\pm0.01$ & -- & -- & -- & -- & $1.6\pm0.1$ & $0.35\pm0.03$ & \\
12090 & & $2.79\pm0.06$ & -- & -- & $0.67\pm0.01$ & -- & -- & -- & -- & $1.6\pm0.1$ & $0.34\pm0.03$ & \\
12072 & & $2.59\pm0.03$ & -- & -- & $0.72\pm0.01$ & -- & -- & -- & -- & $1.5\pm0.1$ & $0.37\pm0.03$ & \\
\hline
\end{tabular}
\tablefoot{We report the parameter values for each models used in the analysis where $k_{\rm B}T$ indicates the temperature, $N_{\rm H}$ the column density, $\Gamma $ the photon index of the power-law model, $\xi$ the ionisation parameter, $v$ the flow velocity (which, in case of the negative number, corresponds to the outflow velocity), $F_{0.5-2\rm\,keV}$ and $F_{2-10\rm\,keV}$ the fluxes in the two different energy ranges. Errors given on parameters are $1\sigma$ errors. We also list the obsID of each observation and the C-statistic and degree of freedom (dof) for every fit. }
\end{sidewaystable*}

%
\begin{table}
\label{tab:lmc_abundance}
 \caption[]{Comparison of element abundances in the Galactic interstellar medium \citep{Lodders10} and in the Large Magellanic Cloud.}
 \centering
\begin{tabular}{lcccc}
 \hline \hline
  $X$  & $A_{\rm Gal}(X)$ & $A_{\rm LMC}(X)$ & $10^{\Delta A(X)}$ & Ref. \\
\hline
He   & 10.987 & 10.94 & 0.897 & (1) \\ 
C    & 8.443  & 8.04  & 0.395 & (1) \\ 
N    & 7.912  & 7.14  & 0.169 & (1) \\ 
O    & 8.782  & 8.04  & 0.181 & (2) \\ 
Ne   & 8.103  & 7.39  & 0.194 & (2) \\ 
Na   & 6.347  & 5.50  & 0.142 & (3) \\ 
Mg   & 7.599  & 6.88  & 0.191 & (2) \\ 
Al   & 6.513  & 5.86  & 0.222 & (4) \\ 
Si   & 7.586  & 6.99  & 0.254 & (2) \\ 
S    & 7.210  & 6.70  & 0.309 & (1) \\ 
Cl   & 5.299  & 4.76  & 0.289 & (1) \\ 
Ar   & 6.553  & 6.29  & 0.546 & (1) \\ 
Ca   & 6.367  & 5.89  & 0.333 & (1) \\ 
Sc   & 3.123  & 2.64  & 0.329 & (1) \\ 
Ti   & 4.949  & 4.81  & 0.678 & (1) \\ 
V    & 4.042  & 4.08  & 1.094 & (1) \\ 
Cr   & 5.703  & 5.47  & 0.585 & (1) \\ 
Mn   & 5.551  & 5.21  & 0.456 & (1) \\ 
Fe   & 7.514  & 6.84  & 0.211 & (1) \\ 
Ni   & 6.276  & 6.04  & 0.581 & (1) \\ 
Zn   & 4.700  & 4.28  & 0.380 & (1) \\ 

\hline
\end{tabular}
\tablebib{(1)~\citet{Russell92}; (2) \citet{Schenck16}; (3)~\citet{Garnett99}; (4) \citet{Korn00}.}
\tablefoot{The third column is the Large Magellanic Cloud abundance relative to the Galactic abundance, which is a parameter of the \texttt{hot} absorption model in \textsc{Spex}. For all the other not listed elements (which are low abundant and hardly contribute to the absorption in the soft X-ray band) the average value $10^{A(X)}=0.4$ is assumed.}
\end{table}

\section{Neutral silicon cross section}
\label{app:neutral_silicon}

\begin{figure}
   \centering
   \includegraphics[width=\hsize]{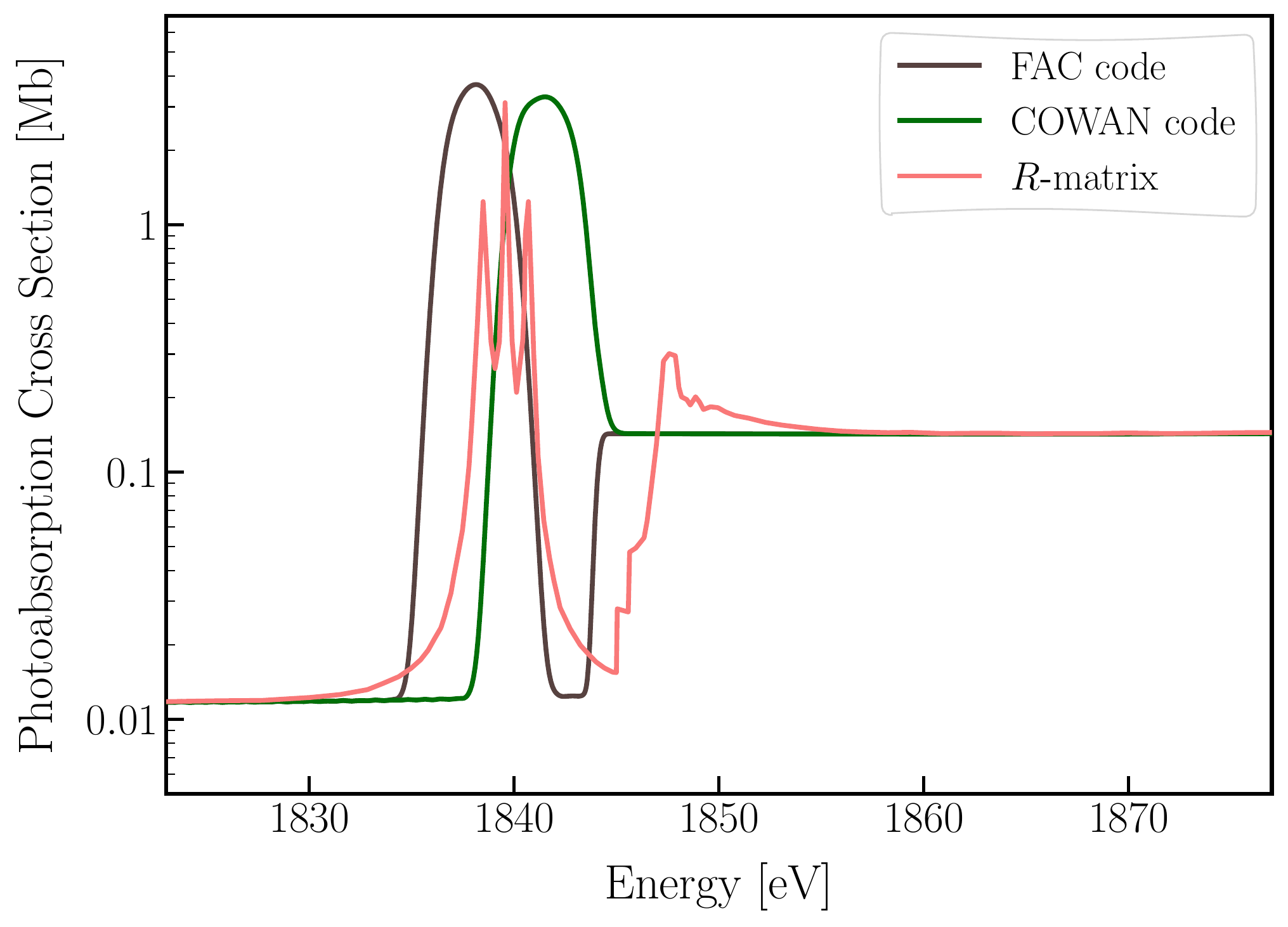}
      \caption{Comparison of the cross-section for neutral silicon, \sii, calculated with different codes. The cross-section calculated with the $R$-matrix, in red, is taken from \cite{Hasoglu18}.
              }
         \label{fig:neutral_silicon}
   \end{figure}

Here we present our calculation of the photoabsorption cross-section for the neutral silicon K-shell in gas form. The inner-shell X-ray absorption for a single, isolated silicon atom implemented in \spex is obtained by independent-particle calculation \citep{Verner93} which returns, a simple step function model where resonant transitions are not included. However, in order to investigate the presence of gaseous atomic silicon in the interstellar medium, in addition to the overwhelming abundance of silicon in cosmic dust, it is important to calculate and include in the model the resonance transition and relaxation effect.\\
In Figure \ref{fig:neutral_silicon} we show the \sii photoabsorption cross-sections obtained from the Flexible Atomic Code \citep[\texttt{FAC},][in black]{Gu08} and the \texttt{COWAN} code \citep[][in blue]{Cowan81}. We overlap for comparison the cross-section calculated by \cite{Hasoglu18} using the modified $R$-matrix method \citep{Berrington95}. The cross-sections show different profiles with a shift in the energy of $1.5-3.5\,\rm eV$. In our analysis, we use the \sii cross-section calculated with the \texttt{COWAN} code since it can represent better the residual in the pre-edge of the Si K-shell for GX 3+1, used as test source in \citetalias{Rogantini19} due to its high signal to noise ratio. 
In Table \ref{tab:fac_cowan} we list the atomic parameters of the \sii transitions, namely the line-centre wavelengths $\lambda_{\rm c}$ and the oscillator strengths $f_{\rm osc}$, necessary to evaluate the optical depth of the lines.

%

\begin{table*}
\label{tab:fac_cowan}
 \caption{Atomic parameter of those \sii transitions which contribute to the line absorption of \sii.}
 \centering
\begin{tabular}{ll@{\hspace{7mm}}ccc@{\hspace{0mm}}c@{\hspace{7mm}}ccc}
 \hline \hline
    &  & \multicolumn{3}{c}{\tt FAC} & & \multicolumn{3}{c}{\tt COWAN} \\\cline{3-5} \cline{7-9}
    \noalign{\vspace{1mm}}
  \multicolumn{1}{c}{\multirow{-2}{*}{Lower energy level}}  & \multicolumn{1}{c}{\multirow{-2}{*}{Upper energy level}} & $\lambda_{\rm c}$ (\AA) & $E_{\rm c}$ (eV) & $f_{\rm osc}$ & & $\lambda_{\rm c}$ (\AA) & $E_{\rm c}$ (eV) & $f_{\rm osc}$ \\
\hline
\noalign{\vspace{1mm}}
     \multirow{4}{*}{$ 1s^2 2s^2 2p^6 3s^2 3p^2\,\, \textsuperscript{3}P_0$} & $ 1s^1 \big(\textsuperscript{4}S\big) 2s^2 2p^6 3s^2 3p^3\,\, \textsuperscript{3}S_1$ & 6.7489 & 1837.1 & $4.19\times10^{-3}$ & & 6.7364 & 1840.5 & $5.36\times10^{-3}$ \\   
      & $ 1s^1 \big(\textsuperscript{2}D\big) 2s^2 2p^6 3s^2 3p^3\,\, \textsuperscript{3}D_1$ & 6.7445 & 1838.3 & $6.10\times10^{-3}$ & & 6.7324 & 1841.6 & $6.64\times10^{-3}$ \\
      & $ 1s^1 \big(\textsuperscript{2}P\big) 2s^2 2p^6 3s^2 3p^3\,\, \textsuperscript{3}P_1$ & 6.7408 & 1839.3 & $3.26\times10^{-3}$ & & 6.7284 & 1842.7 & $3.58\times10^{-3}$ \\
      & $ 1s^1 \big(\textsuperscript{2}P\big) 2s^2 2p^6 3s^2 3p^3\,\, \textsuperscript{1}P_1$ & 6.7393 & 1839.7 & $5.12\times10^{-5}$ & & 6.7273 & 1843.0 & $2.17\times10^{-7}$ \\
     \noalign{\vspace{2mm}}
     \multirow{9}{*}{$ 1s^2 2s^2 2p^6 3s^2 3p^2\,\, \textsuperscript{3}P_1$} & $ 1s^1 \big(\textsuperscript{4}S\big) 2s^2 2p^6 3s^2 3p^3\,\, \textsuperscript{5}S_2$ & 6.7514 & 1836.4 & $3.99\times10^{-7}$ & & 6.7390 & 1839.8 & $4.67\times10^{-7}$ \\  
      & $ 1s^1 \big(\textsuperscript{4}S\big) 2s^2 2p^6 3s^2 3p^3\,\, \textsuperscript{3}S_1$ & 6.7489 & 1837.1 & $1.42\times10^{-2}$ & & 6.7364 & 1840.5 & $1.60\times10^{-2}$ \\ 
      & $ 1s^1 \big(\textsuperscript{2}D\big) 2s^2 2p^6 3s^2 3p^3\,\, \textsuperscript{3}D_1$ & 6.7448 & 1838.2 & $3.73\times10^{-3}$ & & 6.7324 & 1841.6 & $4.75\times10^{-3}$ \\  
      & $ 1s^1 \big(\textsuperscript{2}D\big) 2s^2 2p^6 3s^2 3p^3\,\, \textsuperscript{3}D_2$ & 6.7448 & 1838.2 & $1.32\times10^{-2}$ & & 6.7324 & 1841.6 & $1.50\times10^{-2}$ \\  
      & $ 1s^1 \big(\textsuperscript{2}D\big) 2s^2 2p^6 3s^2 3p^3\,\, \textsuperscript{1}D_2$ & 6.7434 & 1838.6 & $4.27\times10^{-5}$ & & 6.7313 & 1841.9 & $2.65\times10^{-6}$ \\  
      & $ 1s^1 \big(\textsuperscript{2}P\big) 2s^2 2p^6 3s^2 3p^3\,\, \textsuperscript{3}P_0$ & 6.7408 & 1839.3 & $3.37\times10^{-3}$ & & 6.7287 & 1842.6 & $3.81\times10^{-3}$ \\  
      & $ 1s^1 \big(\textsuperscript{2}P\big) 2s^2 2p^6 3s^2 3p^3\,\, \textsuperscript{3}P_1$ & 6.7408 & 1839.3 & $2.52\times10^{-3}$ & & 6.7284 & 1842.7 & $2.88\times10^{-3}$ \\  
      & $ 1s^1 \big(\textsuperscript{2}P\big) 2s^2 2p^6 3s^2 3p^3\,\, \textsuperscript{3}P_2$ & 6.7404 & 1839.4 & $4.09\times10^{-3}$ & & 6.7284 & 1842.7 & $4.39\times10^{-3}$ \\  
      & $ 1s^1 \big(\textsuperscript{2}P\big) 2s^2 2p^6 3s^2 3p^3\,\, \textsuperscript{1}P_1$ & 6.7393 & 1839.7 & $1.91\times10^{-5}$ & & 6.7273 & 1843.0 & $3.93\times10^{-8}$ \\ 
     \noalign{\vskip 2mm}
     \multirow{9}{*}{$ 1s^2 2s^2 2p^6 3s^2 3p^2\,\, \textsuperscript{3}P_2$} & $ 1s^1 \big(\textsuperscript{4}S\big) 2s^2 2p^6 3s^2 3p^3\,\, \textsuperscript{5}S_2$ & 6.7514 & 1836.4 & $3.74\times10^{-6}$ & & 6.7393 & 1839.7 & $1.32\times10^{-6}$ \\  
      & $ 1s^1 \big(\textsuperscript{4}S\big) 2s^2 2p^6 3s^2 3p^3\,\, \textsuperscript{3}S_1$ & 6.7489 & 1837.1 & $2.21\times10^{-2}$ & & 6.7368 & 1840.4 & $2.63\times10^{-2}$ \\ 
      & $ 1s^1 \big(\textsuperscript{2}D\big) 2s^2 2p^6 3s^2 3p^3\,\, \textsuperscript{3}D_1$ & 6.7448 & 1838.2 & $4.57\times10^{-4}$ & & 6.7324 & 1841.6 & $2.80\times10^{-4}$ \\  
      & $ 1s^1 \big(\textsuperscript{2}D\big) 2s^2 2p^6 3s^2 3p^3\,\, \textsuperscript{3}D_2$ & 6.7448 & 1838.2 & $3.99\times10^{-3}$ & & 6.7324 & 1841.6 & $4.49\times10^{-3}$ \\  
      & $ 1s^1 \big(\textsuperscript{2}D\big) 2s^2 2p^6 3s^2 3p^3\,\, \textsuperscript{3}D_3$ & 6.7448 & 1838.2 & $2.44\times10^{-2}$ & & 6.7324 & 1841.6 & $2.72\times10^{-2}$ \\  
      & $ 1s^1 \big(\textsuperscript{2}D\big) 2s^2 2p^6 3s^2 3p^3\,\, \textsuperscript{1}D_2$ & 6.7434 & 1838.6 & $   -             $ & & 6.7313 & 1841.9 & $3.89\times10^{-5}$ \\  
      & $ 1s^1 \big(\textsuperscript{2}P\big) 2s^2 2p^6 3s^2 3p^3\,\, \textsuperscript{3}P_1$ & 6.7408 & 1839.3 & $4.29\times10^{-3}$ & & 6.7284 & 1842.7 & $4.97\times10^{-3}$ \\  
      & $ 1s^1 \big(\textsuperscript{2}P\big) 2s^2 2p^6 3s^2 3p^3\,\, \textsuperscript{3}P_2$ & 6.7404 & 1839.4 & $1.32\times10^{-2}$ & & 6.7284 & 1842.7 & $1.46\times10^{-2}$ \\  
      & $ 1s^1 \big(\textsuperscript{2}P\big) 2s^2 2p^6 3s^2 3p^3\,\, \textsuperscript{1}P_1$ & 6.7393 & 1839.7 & $4.53\times10^{-5}$ & & 6.7273 & 1843.0 & $4.91\times10^{-8}$ \\
      \noalign{\vspace{1mm}}
\hline
\end{tabular}
\tablefoot{$\lambda_{\rm c}$ and $E_{c}$ are the line-centre wavelength and energy, respectively. The oscillator strength, indicated with $f_{\rm osc}$, is dimensionless.}
\end{table*}

\section{Instrumental feature at 6.741 \AA}
\label{app:6.74}
\begin{figure}
   \centering
     \includegraphics[width=\hsize]{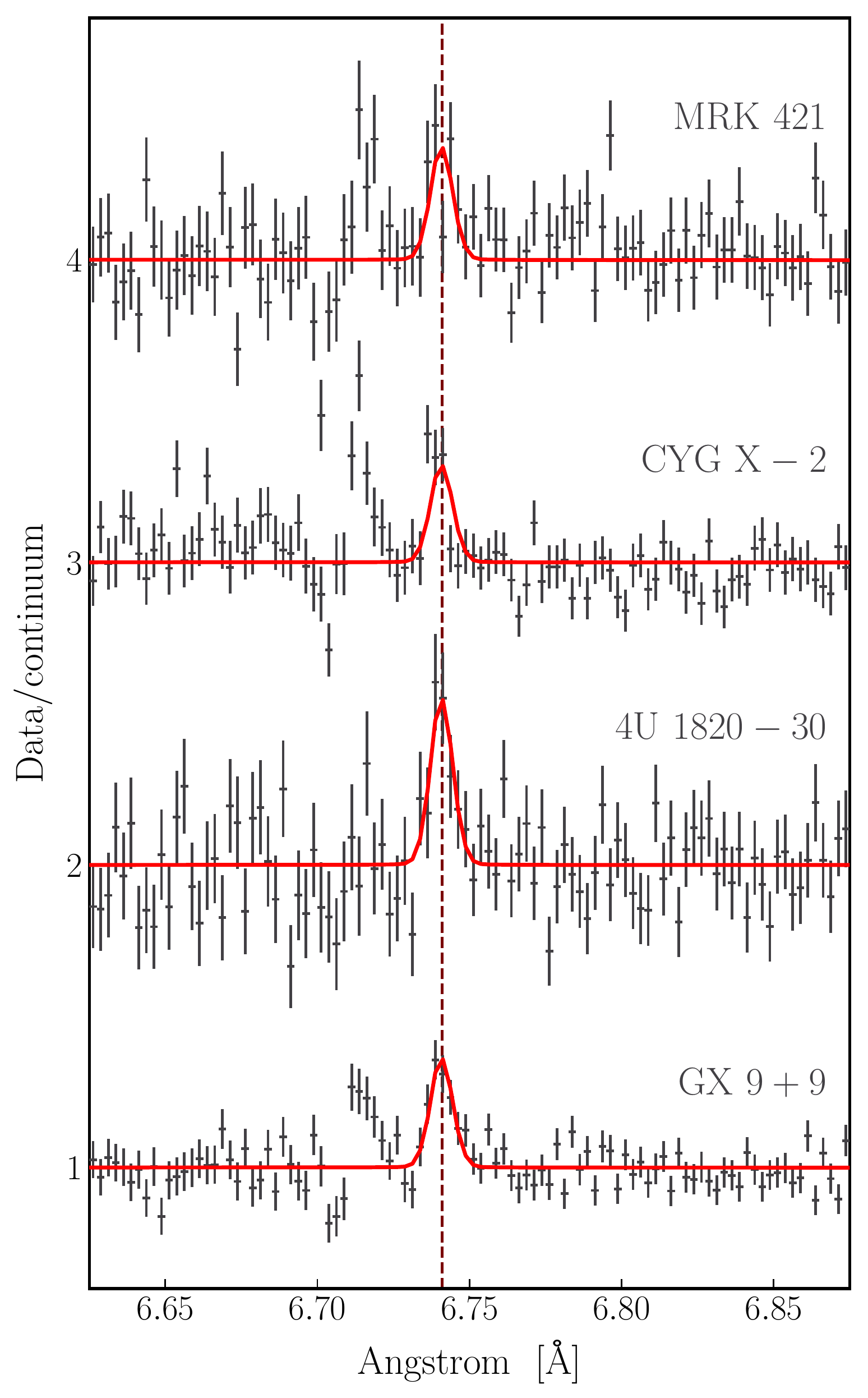}
      \caption{Fit of the line in emission present in the silicon K-edge region. To fit the underlying continuum we use a simple power-law. The different data-sets are shifted along the Y axis for clarity.}
         \label{fig:sik_line}
   \end{figure}

For all the sources in our sample, we observe a line in emission at 6.741 \AA, in the vicinity of the silicon K-edge. This feature was already noticed in other sources, leading to different interpretation (see Section \ref{sec:data_reduction}). In the present sample, we explored first the possibility of a \sixiii forbidden line emission ($\lambda =  6.7405$ \AA). However, this is difficult to explain since, at the ionisation parameter that produces the \sixiii($f$) line, we expect to detect also the \nex and \mgxii lines, which are not observed. Their ionic densities would all peak at the same ionisation parameter \citep[Figure 6 of][]{Mehdipour16}.\\ 
The emission feature could be also associated with the scattering peak of Si. This possibility was explored in \citetalias{Rogantini19}. However, this apparent emission feature was also observed in sources that should not display either an emission line, as they present a featureless spectrum, or a scattering peak, as their column density is too low to produce absorption by Si (e.g., Mrk 421).\\
Here we explore the possibility of an instrumental line. We consider the Galactic X-ray binaries, \object{GX 9+9}, \object{4U 1820-30}, \object{Cyg X-2}, and the blazar \object{MRK 421}. In Figure \ref{fig:sik_line}, we show the silicon K-edge region of the previous sources subtracted by the underlying continuum. We fit simultaneously the line with a delta function and we find a line-centre wavelength of $\lambda_{\rm c}= 6.741\pm0.001$.
Moreover, we analyse separately the $\pm 1$ order of MEG and HEG for sources (e.g. 4U 1636-53) with a relatively lower flux, to minimize the pileup effect. We notice that only $+1$ MEG, which does not cross the front-illuminated chip in the silicon region, does not display the emission peak. We suggest therefore that the deep calibration silicon feature present in the effective area of $-1$ MEG and $\pm 1$ HEG could play a role in displaying the spike at 6.741 \AA. 

\end{appendix}

\end{document}